\documentclass[preprint,3p,12pt]{elsarticle}
\usepackage{mathrsfs}
\usepackage{amsmath}
\usepackage{amsthm}
\usepackage{stmaryrd}
\usepackage{bbding}
\usepackage{dcolumn}
\usepackage{graphicx}
\usepackage{amsfonts}
\usepackage{amssymb}
\usepackage{psfrag}
\usepackage{wrapfig}
\usepackage{subfigure}
\usepackage{makeidx}
\usepackage{bm}
\usepackage{epsf}
\usepackage{epsfig}
\usepackage{setspace}
\usepackage{graphicx}
\usepackage{epstopdf}
\usepackage{psfrag}
\usepackage{subfigure}
\usepackage{color}

\usepackage[ruled,linesnumbered]{algorithm2e}

\definecolor{customgreen}{rgb}{0,0.6,0}

\begin{document}

\title{Efficient Wall-Modeled High-Order Compact Gas-Kinetic Scheme for Compressible Turbulent Flows}

\author[HKUST1]{Yaqing Yang}
\ead{yangyq@ust.hk}  

\author[HKUST1]{Fengxiang Zhao}
\ead{fzhaoac@connect.ust.hk}

\author[HKUST1,HKUST2]{Kun Xu \corref{cor}}
\ead{makxu@ust.hk}

\cortext[cor]{Corresponding author}
\address[HKUST1]{Department of Mathematics, Hong Kong University of Science and Technology, Clear Water Bay, Kowloon, Hong Kong}
\address[HKUST2]{Shenzhen Research Institute, Hong Kong University of Science and Technology, Shenzhen, China}
 
\begin{abstract}

Scale-resolving simulations of wall-bounded turbulent flows remain prohibitively expensive at high Reynolds numbers, owing to the stringent near-wall resolution requirements. High-order compact gas-kinetic schemes (CGKS) are accurate, robust, and efficient for compressible flows, making them an attractive foundation for reducing this cost. Building on the fifth-order scheme CGKS-5th, we develop a wall-modeled CGKS framework that alleviates the near-wall resolution burden through a pressure-gradient-based non-equilibrium wall model while preserving the resolving power of the outer solver.
CGKS-5th resolves the outer flow and supplies the wall model with data at the exchange location. On coarse near-wall meshes, the wall model reconstructs the under-resolved viscous wall stress, while CGKS-5th provides the inviscid wall flux directly; the two combine to form the wall momentum flux. To capture non-equilibrium effects in adverse-pressure-gradient and separated regions, the wall model retains a pressure-gradient source term together with a pressure-gradient-corrected near-wall damping function. 
We assess the framework on two distinct flows: bluff-body separation past a circular cylinder, and a shock-induced separation bubble on the transonic RAE 2822 airfoil, using near-wall meshes far coarser than wall-resolved simulations require. For the RAE 2822 case, this corresponds to a roughly twentyfold coarsening in the wall-normal direction, with comparable coarsening in the streamwise and spanwise directions. In both cases, the wall-modeled CGKS-5th reproduces the separated flow structures and markedly improves near-wall predictions over its wall-model-free counterpart, most notably the skin-friction coefficient. The framework thus delivers accurate predictions of these separated flows at substantially reduced near-wall cost, while its lightweight coupling adds less than 1\% runtime overhead in a multi-GPU implementation.

\end{abstract}
 

\maketitle
 
\section{Introduction}

Turbulent flows at high Reynolds numbers are ubiquitous in engineering applications such as aerospace and turbomachinery. Among these flows, large-scale separation behind bluff bodies and shock-induced separation caused by shock-wave/boundary-layer interaction (SWBLI) represent two common and challenging phenomena, typically characterized by strong unsteadiness, pronounced adverse pressure gradients, and complex separation-reattachment structures.
As the Reynolds number increases, the viscous near-wall scales become increasingly small, rendering direct numerical simulation (DNS) and wall-resolved large-eddy simulation (WRLES) prohibitively expensive at conditions of engineering interest.
Achieving high accuracy at an affordable cost is therefore a central challenge in applying scale-resolving simulation methods to practical engineering flows.

Compared with conventional second-order methods, high-order schemes offer lower numerical dissipation and higher resolution on a given mesh, and are
therefore attractive for reducing the computational cost required to achieve a target accuracy. Considerable effort has been devoted to their development,
including high-order finite-difference (FD) methods \cite{FD-1,FD-2}, the essentially non-oscillatory (ENO) scheme \cite{ENO-1,ENO-2}, the weighted
essentially non-oscillatory (WENO) scheme \cite{WENO-1,WENO-2}, the Hermite WENO (HWENO) scheme \cite{HWENO-1}, the discontinuous Galerkin (DG) method
\cite{DG-0,DG-1}, the flux reconstruction (FR) approach \cite{FR}, and the correction procedure via reconstruction (CPR) \cite{CPR}, among others.

In recent years, high-order compact gas-kinetic schemes (CGKS) have been developed for the high-fidelity simulation of compressible flows \cite{GKS-high-1, CGKS-high-2, CGKS-high-3, CGKS-high-4}. These schemes build upon the gas distribution function of the underlying gas-kinetic scheme (GKS) framework \cite{GKS-Xu1, GKS-Xu2, GKS-Xu3}, endowing the scheme with several intrinsic advantages: the inviscid and viscous fluxes are evaluated simultaneously within a single unified formulation; the flux transitions naturally between an equilibrium form in smooth regions and a non-equilibrium form near discontinuities; and second-order temporal accuracy can be achieved through a single-step evolution. Building on this framework, we proposed in our previous work a fifth-order compact gas-kinetic scheme (CGKS-5th) and validated it on a series of turbulent benchmark cases, confirming its accuracy and robustness for complex compressible flows \cite{CGKS-5th-novel}. It should be emphasized, however, that although high-order schemes relax the overall mesh-resolution constraints by improving the accuracy attainable on a given mesh, they still impose stringent resolution requirements in the near-wall region. This requirement, together with the cost of long-time simulation, remains the principal obstacle to applying CGKS-5th to realistic engineering flows. A promising route to relieving this near-wall burden is wall modeling.

Wall modeling is a general near-wall treatment adopted across a wide range of turbulence simulation frameworks, from wall functions in Reynolds-averaged
Navier--Stokes (RANS) methods \cite{WM-RANS} to hybrid RANS--LES methods such as detached-eddy simulation (DES) \cite{WM-DES} and LES-type approaches
\cite{WM-LES-1, WM-LES-2}. According to their underlying assumptions, wall models are commonly classified as equilibrium or non-equilibrium. Equilibrium models neglect the pressure-gradient and convective terms, and are implemented either through an algebraic law-of-the-wall relation \cite{WM-EQA-1,WM-EQA-2} or by solving a simplified near-wall ordinary differential equation (ODE) \cite{WM-EQ-1, WM-EQ-2}. Their simplicity and low cost make them widely used, but their validity hinges on the assumptions of local equilibrium and negligible pressure gradient. These assumptions break down in flows with separation, reattachment, strong adverse pressure gradients, or SWBLI, where equilibrium models may lead to substantial errors in the wall shear stress. Non-equilibrium models address this limitation by retaining, to different extents, pressure-gradient and convective effects \cite{WM-NEQ-1}, thereby accounting for departures from equilibrium. Furthermore, because the friction velocity alone may not characterize the near-wall profile in such flows, introducing a pressure-gradient-corrected velocity scale can further improve the representation of adverse-pressure-gradient and separated boundary layers \cite{WM-NEQ-2}.

Despite the maturity of wall modeling across RANS and LES frameworks, its application within CGKS remains largely unexplored. Building on CGKS-5th, we develop a wall-modeled CGKS framework equipped with a non-equilibrium wall model. To capture non-equilibrium effects, the wall model retains the pressure-gradient term in the near-wall ODE and adopts a pressure-gradient-corrected velocity scale in the near-wall damping function. The coupling is achieved through a momentum-flux boundary condition, in which the wall model corrects only the viscous part of the wall-cell flux while CGKS-5th provides the inviscid part. 
Two representative three-dimensional separated flows are selected for validation: subsonic flow over a circular cylinder, featuring laminar separation and turbulent wake, and high-Reynolds-number transonic flow over the RAE 2822 airfoil, involving SWBLI accompanied by a separation bubble. The results demonstrate that, on coarsened near-wall meshes, the proposed framework improves the prediction of wall quantities, in particular the skin-friction coefficient, while reliably capturing the main separation features, at substantially reduced near-wall cost and with negligible coupling overhead.

The remainder of this paper is organized as follows. Section 2 introduces the gas-kinetic scheme and the finite-volume framework. Section 3 briefly reviews
the fifth-order compact gas-kinetic scheme used as the outer solver. Section 4 describes the non-equilibrium wall model and its implementation
within CGKS-5th. Numerical examples are provided in Section 5, and conclusions are drawn in Section 6.

\section{Gas-kinetic scheme and finite volume method}

In the gas-kinetic method, time-dependent numerical fluxes are constructed from a space-time coupled solution of the BGK equation for the gas distribution function \cite{GKS-Xu1,GKS-Xu2}.
The BGK equation is a simplification of the Boltzmann equation, and the three-dimensional BGK equation \cite{BGK-1,BGK-2} can be written as
\begin{equation}\label{bgk}
f_t+\boldsymbol{u}\cdot \nabla f=\frac{g-f}{\tau},
\end{equation}
where $\boldsymbol{u}=(u,v,w)$ is the particle velocity, $\tau$ is the collision time and $f$ is the gas distribution function. 
The equilibrium state $g$ is given by Maxwellian distribution 
\begin{equation*}
g=\rho\left(\frac{\lambda}{\pi}\right)^{(N+3)/2}e^{-\lambda[(\boldsymbol{u}-\boldsymbol{U})^2+\boldsymbol{\xi}^2]},
\end{equation*}
where $\rho$ is the density, $\boldsymbol{U}=(U,V,W)$ is the macroscopic fluid velocity, and $\lambda=\displaystyle\frac{\rho}{2p}$, $p$ is the pressure. 
In the BGK equation, the collision operator involves a simple relaxation from $f$ to the local equilibrium state $g$. 
The variable $\boldsymbol{\xi}$ accounts for the internal degree of freedom of molecular motion, $\boldsymbol{\xi}^2=\xi_1^2+\dots+\xi_N^2$, where $N=(5-3\gamma)/(\gamma-1)$ is the number of internal degree of freedom and $\gamma$ is the specific heat ratio. 
The collision term satisfies the compatibility condition
\begin{equation*}
\int \frac{g-f}{\tau}\boldsymbol{\psi} \, \text{d}\Xi=0,
\end{equation*}
where $\displaystyle\boldsymbol{\psi}=\left(1,u,v,w,\frac{1}{2}(u^2+v^2+w^2+\boldsymbol{\xi}^2)\right)^T$ and $\text{d}\Xi=\text{d}u\text{d}v\text{d}w\text{d}\xi_1\dots\text{d}\xi_{N}$.
The macroscopic conservative variables $\boldsymbol{Q}=(\rho, \rho U,\rho V, \rho W, \rho E)$ can be calculated through the gas distribution function $f$
\begin{equation}\label{Q-macro}
\boldsymbol{Q}=\int f \boldsymbol{\psi} \,\text{d}\Xi,
\end{equation}
and the corresponding fluxes can be given by taking moments of the gas
distribution function
\begin{align}\label{flux-macro}
\boldsymbol{F}=\int \boldsymbol{u} f \boldsymbol{\psi} \, \text{d}\Xi,
\end{align}
where $\boldsymbol{F}$ represents either the Euler flux or the Navier-Stokes (NS) flux, depending on the order of approximation of $f$ to $g$. 
Based on the integral solution of BGK equation Eq.\eqref{bgk}, the gas distribution function $f(\boldsymbol{x}_{G},t,\boldsymbol{u},\boldsymbol{\xi})$ can be given by
\begin{equation*}
f(\boldsymbol{x}_{G},t,\boldsymbol{u},\boldsymbol{\xi})=\frac{1}{\tau}\int_0^t
g(\boldsymbol{x}',t',\boldsymbol{u}, \boldsymbol{\xi})e^{-(t-t')/\tau}\text{d}t'+e^{-t/\tau}f_0(-\boldsymbol{u}t,\boldsymbol{\xi}).
\end{equation*}
By modeling the unknown terms in the integral solution, a second-order explicit gas distribution function is obtained as
\begin{align}\label{flux}
f(\boldsymbol{x}_{G},t,\boldsymbol{u},\boldsymbol{\xi})=&(1-e^{-t/\tau})g_0+
[(t+\tau)e^{-t/\tau}-\tau](\overline{a}_1u+\overline{a}_2v+\overline{a}_3w)g_0\nonumber\\
+&(t-\tau+\tau e^{-t/\tau}){\bar{A}} g_0\nonumber\\
+&e^{-t/\tau}g_r[1-(\tau+t)(a_{1}^{r}u+a_{2}^{r}v+a_{3}^{r}w)-\tau A^r](1-H(u))\nonumber\\
+&e^{-t/\tau}g_l[1-(\tau+t)(a_{1}^{l}u+a_{2}^{l}v+a_{3}^{l}w)-\tau A^l]H(u),
\end{align}
where $g_{0}$, $g_l$ and $g_r$ are determined by the conservative variables reconstructed at the interface, and the coefficients are determined by the macroscopic variables and their derivatives. The specific formulas can be referred to in \cite{GKS-Xu1,GKS-Xu2}.

For the second-order evolution solution, the time accurate distribution function $f(t)$ at cell interface can be approximated through a linearization in time \cite{CGKS-high-5}
\begin{equation*}
\hat{f}(t)=f^n+t f_t^n,
\end{equation*}
where
\begin{align*}
    f^n&=\big(4\bar{f}(\Delta t/2) - \bar{f}(\Delta t)\big)/\Delta t,\\
    f_t^n&=4\big(\bar{f}(\Delta t) - 2\bar{f}(\Delta t/2)\big)/{\Delta t}^2,
\end{align*}
$\bar{f}(\Delta t)$ and $\bar{f}(\Delta t/2)$ are the time integrations of $f(t)$ over the interval $[t_n, t_n + \Delta t]$ and $[t_n, t_n + \Delta t/2]$, respectively.
The numerical fluxes and their time derivatives can be obtained by taking moments of $\hat{f}(t)$ and $\hat{f}_t(t)$ at $t = t_n$
\begin{equation*}
\boldsymbol{F}^n = \int \boldsymbol{u} f^n \boldsymbol{\psi} \, \text{d}\Xi,\quad
\boldsymbol{F}_t^n = \int \boldsymbol{u} f_t^n \boldsymbol{\psi} \, \text{d}\Xi.
\end{equation*}
Simultaneously, the flow variables and their time derivatives can be given by
\begin{equation*}
\boldsymbol{Q}^n = \int f^n \boldsymbol{\psi} \, \text{d}\Xi,\quad
\boldsymbol{Q}_t^n = \int f_t^n \boldsymbol{\psi} \, \text{d}\Xi.
\end{equation*}
The obtained conservative variables at interfaces are used to achieve compact reconstructions in high-order compact GKS.

Taking moments of Eq.~\eqref{bgk} and integrating with respect to space, the semi-discretized finite volume scheme can be obtained as
\begin{align}\label{semi}
\frac{\text{d} \boldsymbol{Q}_i}{\text{d} t}=\mathcal{L}_{i}(\boldsymbol{Q}),
\end{align}
where $\boldsymbol{Q}_i$ is the cell-averaged conservative variables over cell $\Omega_{i}$. 
The operator $\mathcal{L}_{i}$ is defined as
\begin{equation*}
\mathcal{L}_{i}(\boldsymbol{Q})
=-\frac{1}{|\Omega_{i}|} \int_{\partial\Omega_{i}}\boldsymbol{F}(t)\boldsymbol{n}\,\text{d}S,
\end{equation*}
where $|\Omega_{i}|$ is the volume of $\Omega_{i}$, $\partial\Omega_{i}$ represents the cell interfaces of $\Omega_{i}$, and $\boldsymbol{n}$ is the unit direction of the cell interface. 
To achieve the expected order of accuracy, the Gaussian 
quadrature is used for the flux integration
\begin{align*}
\int_{\sigma_{j}}\boldsymbol{F}(t)\boldsymbol{n}\,\text{d}S=\sum_{k=1}^{K}\omega_{k}\boldsymbol{F}(\boldsymbol{x}_k,t)S_{j},
\end{align*}
where $S_{j}$ is the area of $\sigma_{j}$, $\sigma_{j}$ is one of the cell interfaces, $\omega_{k}$ is the Gaussian quadrature weight, $\boldsymbol{x}_{k}$ is the Gaussian quadrature point, $K$ is the total number of Gaussian quadrature points.
The numerical flux $\boldsymbol{F}(\boldsymbol{x}_k,t)$ at Gaussian quadrature point can be given by Eq.\eqref{flux-macro}.
In this study, each face is a quadrilateral with $K=4$ Gaussian points for high-order schemes.

\section{High-order compact gas-kinetic scheme}

The time-dependent gas distribution function in GKS provides not only the numerical fluxes but also the pointwise conservative variables at cell interfaces, offering a natural framework for high-order compact extensions. Recently, a new fifth-order compact GKS (CGKS-5th) on three-dimensional structured meshes has been developed \cite{CGKS-5th-novel}, achieving high accuracy, efficiency, and resolution. The scheme has been validated on challenging numerical test cases for compressible flows, and its formulation is briefly introduced below.

\subsection{Flow variable update and temporal discretization}

For CGKS-5th, the cell-averaged conservative variable $\boldsymbol{Q}(t)$ within a control volume $\Omega$ is updated according to the conservation laws, as given in Eq.\eqref{semi}. The cell-averaged derivatives of $\boldsymbol{Q}(t)$ over $\Omega$ are updated using Gauss's theorem
\begin{equation*}\label{derivative}
\displaystyle\nabla \boldsymbol{Q}(t)=\frac{1}{|\Omega|}\int_{\Omega}\nabla \boldsymbol{Q}(t)\text{d}V=\frac{1}{|\Omega|}\int_{\partial\Omega}\boldsymbol{Q}(t) \boldsymbol{n}\text{d}S,
\end{equation*}
where $\boldsymbol{n}$ is the unit normal vector on the cell interface.
Additionally, the line-averaged partial derivative along a segment defined by points $\boldsymbol{x}_{1}$ and $\boldsymbol{x}_{2}$ is given by
\begin{equation*}
(\partial_{l} \boldsymbol{Q})(t) = \frac{1}{\left|l_G\right|}\int_{\boldsymbol{x}_{1}}^{\boldsymbol{x}_{2}}\frac{\partial \boldsymbol{Q}(t)}{\partial l}\text{d}l = \frac{1}{\left|l_G\right|}\left(\boldsymbol{Q}(\boldsymbol{x}_{2},t)-\boldsymbol{Q}(\boldsymbol{x}_{1},t)\right),
\end{equation*}
where $\left|l_G\right| = \left|\boldsymbol{x}{2}-\boldsymbol{x}{1}\right|$, $\boldsymbol{Q}(\boldsymbol{x}_{1},t)$ and $\boldsymbol{Q}(\boldsymbol{x}_{2},t)$ are the values of $\boldsymbol{Q}$ at $\boldsymbol{x}_{1}$ and $\boldsymbol{x}_{2}$ inside the cell, respectively. At each time step, $\boldsymbol{Q}(\boldsymbol{x}_{1},t)$ and $\boldsymbol{Q}(\boldsymbol{x}_{2},t)$ are obtained from the time-accurate solution of the gas distribution function in the GKS. 

According to Eq.\eqref{semi}, the fully discrete form of the conservation laws over cell $\Omega_{i}$ can be represented as
\begin{equation*}
\boldsymbol{Q}_{i}^{n+1}=\boldsymbol{Q}_{i}^{n}+\int_{t^n}^{t^{n+1}} \mathcal{L}(\boldsymbol{Q}_{i},t)\,\mathrm{d}t,
\end{equation*}
where $\boldsymbol{Q}_{i}^{n+1}$ represents the cell-averaged conservative variable over cell $\Omega_{i}$ at $t_{n+1}=t_n+\Delta t$.
In the CGKS-5th, the cell-averaged conservative variables are updated with high-order temporal accuracy based on a two-stage, fourth-order method \cite{s2o4-0,GRP-high-1,GKS-high-1}
\begin{align*}
\boldsymbol{Q}_{i}^{*}&=\boldsymbol{Q}_{i}^n+\frac{1}{2}\Delta t\mathcal{L}(\boldsymbol{Q}_{i}^n)+\frac{1}{8}\Delta t^2\frac{\partial}{\partial t}\mathcal{L}(\boldsymbol{Q}^n_{i}), \\
\boldsymbol{Q}_{i}^{n+1}&=\boldsymbol{Q}_{i}^n+\Delta t\mathcal {L}(\boldsymbol{Q}_{i}^n)+\frac{1}{2}\Delta t^2\frac{\partial}{\partial t}\mathcal{L}(\boldsymbol{Q}^n_{i})-\frac{1}{3}\Delta t^2\frac{\partial}{\partial t}\widetilde{\mathcal{L}}(\boldsymbol{Q}^n_{i})+\frac{1}{3}\frac{\partial}{\partial t}\widetilde{\mathcal{L}}(\boldsymbol{Q}^{*}_{i}),
\end{align*}
where $\widetilde{\mathcal{L}}$ is a nonlinear operator \cite{GKS-high-ns2o4}. In this work, the nonlinear limiter employed in $\widetilde{\mathcal{L}}$ is determined directly from the spatial reconstruction without introducing any additional computation.
Simultaneously, the cell-averaged gradients and line-averaged derivatives are evaluated from the interface conservative variables, with the variables on each side of the interface evolved via a two-stage method for third-order accuracy
\begin{align*}
\boldsymbol{Q}^{*}&=\boldsymbol{Q}^n+\frac{1}{2}\Delta t (\partial_t \boldsymbol{Q})^n,\\
\boldsymbol{Q}^{n+1}&=\boldsymbol{Q}^n+\Delta t (\partial_t \boldsymbol{Q})^{*}.
\end{align*}
Since the conservative variables on each side of an interface may differ near discontinuities \cite{CGKS-high-5,CGKS-high-4}, the update strategy for $\boldsymbol{Q}(t)$ is formulated as
\begin{align*}
\boldsymbol{Q}^l(t)&=(1-e^{-\Delta t /\tau_0})\boldsymbol{Q}^e(t)+e^{-\Delta t /\tau_0}\boldsymbol{Q}_0^l(t),\\
\boldsymbol{Q}^r(t)&=(1-e^{-\Delta t /\tau_0})\boldsymbol{Q}^e(t)+e^{-\Delta t /\tau_0}\boldsymbol{Q}_0^r(t).
\end{align*}
The values and time derivatives of the conservative variables on the left and right sides of the interface are obtained from this formula. Here, $\boldsymbol{Q}_0^l$, $\boldsymbol{Q}_0^r$, and $\boldsymbol{Q}^e$ are respectively given by the GKS distribution function at the interface.
    
\subsection{Fifth-order compact gas-kinetic scheme}

For the target cell $\Omega_0$, its interfaces are denoted by $F_m$ ($m = 1, \dots, 6$), and its edges are denoted by $E_n$ ($n = 1, \dots, 12$). 
The cell adjacent to $\Omega_0$ that shares the face $F_m$ is represented by $\Omega_{1,m}$, while the cell adjacent to $\Omega_0$ that shares only the edge $E_n$ is represented by $\Omega_{2,n}$.

To achieve fifth-order accuracy, a compact stencil containing all the face-neighboring and edge-neighboring cells is defined as follows
\begin{equation*}
S^{cell-5th}=\{\Omega_0, \Omega_{1,m}, \Omega_{2,n}\,|\,m = 1, \dots, 6,\, n = 1, \dots, 12\}.
\end{equation*}
The fifth-order reconstruction employs the following data
\begin{align*}\label{stencil-5}
S^{data-5th}=S^{0}\cup S^{face}\cup S^{edge},
\end{align*}
where
\begin{align*}
    S^{0}&=\{\boldsymbol{Q}_0,\,(\partial_l \boldsymbol{Q})_{0,G}\},\\
    S^{face}&=\{\boldsymbol{Q}_{1,m},\,(\nabla \boldsymbol{Q})_{1,m}\,|\,m = 1, \dots, 6\},\\
    S^{edge}&=\{\boldsymbol{Q}_{2,n},\,(\partial_{\boldsymbol{\tau}} \boldsymbol{Q})_{2,n}\,|\,n=1,\dots,12\}.
\end{align*}
$\boldsymbol{Q}_0$, $\boldsymbol{Q}_{1,m}$ and $\boldsymbol{Q}_{2,n}$ represent the cell-averaged conservative values over cells $\Omega_0$, $\Omega_{1,m}$ and $\Omega_{2,n}$, respectively. $(\partial_{l} \boldsymbol{Q})_{0,G}$ represents the line-averaged partial derivatives within the target cell $\Omega_0$. 
$(\nabla \boldsymbol{Q})_{1,m}$ denotes the cell-averaged gradient over cell $\Omega_{1,m}$, where $\nabla=(\partial_x,\partial_y,\partial_z)$. 
$(\partial_{\boldsymbol{\tau}} \boldsymbol{Q})_{2,n}$ denotes the cell-averaged directional derivative over the cell $\Omega_{2,n}$, with $\partial_{\boldsymbol{\tau}} = (\partial_{\tau_1}, \partial_{\tau_2})$, where $\partial_{\tau_1}$ and $\partial_{\tau_2}$ are linear combinations of $\partial_x$, $\partial_y$ and $\partial_z$. Details of computing these quantities are given in \cite{CGKS-5th-novel}.

From the stencil $S^{data-5th}$, a quartic polynomial $P^4(\boldsymbol{x})$ is reconstructed for the target cell $\Omega_0$ as
\begin{equation*}
P^4(\boldsymbol{x})=\boldsymbol{Q}_{0}+\sum_{|\boldsymbol{d}|=1}^4 a_{\boldsymbol{d}}p_{\boldsymbol{d}}(\boldsymbol{x}),
\end{equation*}
where $\boldsymbol{Q}_{0}$ is the cell-averaged variable over cell $\Omega_{0}$, the multi-index $\boldsymbol{d}=(d_1, d_2, d_3)$, and $|\boldsymbol{d}|=d_1+d_2+d_3$, and $p_{\boldsymbol{d}}(\boldsymbol{x})$ is the zero-mean basis. 
To determine the polynomial $P^4(\boldsymbol{x})$, the following system is solved using the constrained least-squares method
\begin{equation}\label{compact-big-5}
\begin{split}
\frac{1}{\left|\Omega_{k}\right|}\int_{\Omega_{k}}P^4(\boldsymbol{x})\text{d}V&=\boldsymbol{Q}_{k},~\Omega_{k}\in S^{cell-5th},\\
\frac{h_1}{\left|\Omega_{1,m}\right|}\int_{\Omega_{1,m}}\nabla P^4(\boldsymbol{x})\text{d}V&=h_1\cdot(\nabla \boldsymbol{Q})_{1,m},~m=1,\dots, 6,\\
\frac{h_2}{\left|\Omega_{2,n}\right|}\int_{\Omega_{2,n}}\frac{\partial}{\partial_{\boldsymbol{\tau}}}P^4(\boldsymbol{x})\text{d}V&=h_2\cdot(\partial_{\boldsymbol{\tau}} \boldsymbol{Q})_{2,n},~n=1,\dots,12,\\
\frac{h_3}{\left|l_G\right|}\int_{l_G}\frac{\partial}{\partial_{l}}P^4(\boldsymbol{x})\text{d}l&=h_3\cdot(\partial_{l} \boldsymbol{Q})_{0,G},
\end{split}
\end{equation}
where the parameters $h_1$, $h_2$, and $h_3$ are scaling factors that improve the conditioning of the reconstruction matrix, whose specific values are given in \cite{CGKS-5th-novel}. 

\subsection{Nonlinear fifth-order compact gas-kinetic scheme}

To handle discontinuities, the generalized ENO (GENO) nonlinear reconstruction \cite{GENO} is adopted, which is dominated by a second-order reconstruction with the ENO property at discontinuities. For this second-order reconstruction, we adopt a more robust sub-stencil strategy than that used in \cite{CGKS-5th-novel}. The sub-stencils are
\begin{equation*}
S_{m}^{cell-2}=\{\Omega_0, \Omega_{1,m}\},
\end{equation*}
and the linear polynomial on each sub-stencil is constructed from
\begin{align}\label{stencil-2}
S_{m}^{data-2}=\{\boldsymbol{Q}_0,\,\boldsymbol{Q}_{1,m},\,(\nabla \boldsymbol{Q})_{1,m}\},\,m=1,\dots,6,
\end{align}
where $\boldsymbol{Q}_{1,m}$ and $(\nabla \boldsymbol{Q})_{1,m}$ are the cell-averaged conservative value and gradient over cell $\Omega_{1,m}$. 
From $S_{m}^{data-2}$ ($m=1,\dots,6$), linear polynomials $P_m^1(\boldsymbol{x})$ are constructed as
\begin{equation}\label{p1}
P_m(\boldsymbol{x})=\boldsymbol{Q}_{0}+\sum_{|\boldsymbol d|=1}b_{\boldsymbol d}^mp_{\boldsymbol d}(\boldsymbol{x}).
\end{equation}
To determine these linear polynomials, the following system is solved using a constrained least-squares method
\begin{equation}\label{compact-sub-2}
\begin{split}
\frac{1}{\left|\Omega_{1,m}\right|}\int_{\Omega_{1,m}}P_m^1(\boldsymbol{x})\text{d}V&=\boldsymbol{Q}_{1,m},\\
\frac{1}{\left|\Omega_{1,m}\right|}\int_{\Omega_{1,m}}\nabla P_m^1(\boldsymbol{x})\text{d}V&=(\nabla \boldsymbol{Q})_{1,m}.
\end{split}
\end{equation}

With the reconstructed polynomials $P^4(\boldsymbol{x})$ and $P_m^1(\boldsymbol{x}), m=1,...,6$, GENO is applied to achieve nonlinear reconstruction, adaptively transitioning from high-order accuracy in smooth regions to second-order near discontinuities to balance resolution and robustness.
The conservative value $\boldsymbol{Q}(\boldsymbol{x}_{G})$ and the spatial derivatives $\partial_{x,y,z} \boldsymbol{Q}(\boldsymbol{x}_{G})$ at Gaussian quadrature point are computed as
\begin{equation}\label{weno-new}
\begin{split}
\boldsymbol{Q}(\boldsymbol{x}_{G})&=\chi P^4(\boldsymbol{x}_{G}) + (1-\chi)\left(\sum_{m=1}^{6}\omega_{m} P_m^1(\boldsymbol{x})\right),\\
\partial_{x,y,z} \boldsymbol{Q}(\boldsymbol{x}_{G})&=\chi \partial_{x,y,z} P^4(\boldsymbol{x}_{G}) + (1-\chi)\left(\sum_{m=1}^{6}\omega_{m} \partial_{x,y,z} P_m^1(\boldsymbol{x})\right),
\end{split}
\end{equation}
where 
\begin{equation*}
\displaystyle\omega_{m}=\frac{\overline{\omega_{m}}}{\sum_{m=1}^{6} \overline{\omega_{m}}},\,
\displaystyle\overline{\omega_{m}}=\frac{d_m}{(IS_m+\epsilon)^5},\,
d_1=\dots=d_6=\displaystyle\frac{1}{6}.
\end{equation*}
Details of the above computation can be found in \cite{GENO, CGKS-5th-novel}.

To accommodate non-orthogonal meshes with low memory usage and simple implementation, the CGKS-5th scheme maps each target cell to a single reference cell \cite{CGKS-5th-novel}, so that all reconstructions are expressed and stored in a unified polynomial form on the reference cell. Reconstructed values are then mapped back to physical coordinates through the inverse transformation.

\section{Wall-modeled high-order compact gas-kinetic scheme}
\label{sec:wall_model}

While the CGKS-5th achieves high accuracy and efficiency, directly resolving the near-wall turbulent structures of complex wall-bounded flows remains computationally prohibitive, as it requires an excessively fine grid in the near-wall region. To relax this requirement, we develop a wall-modeled CGKS framework in which the near-wall layer is represented by a wall model rather than fully resolved.

\subsection{Non-equilibrium wall-model formulation}
\label{subsec:formulation}

A wide range of wall models, from equilibrium models to those based on the full boundary-layer equations, have been extensively discussed in the literature \cite{WM-LES-2}. The present work employs a non-equilibrium wall model that incorporates the wall-parallel pressure gradient. To implement this model, a one-dimensional embedded mesh is constructed along the wall-normal direction at each wall interface, extending from the wall to an exchange location. A simplified momentum equation is solved on this mesh, with the outer CGKS-5th solution at the exchange location supplying the boundary condition. The resulting wall shear stress is then returned to the outer solver as the wall momentum flux.

As a first step toward a non-equilibrium wall-modeled CGKS framework, we focus on flows whose departure from equilibrium is governed primarily by the wall-parallel pressure gradient, for which the convective and unsteady terms play a minor role in the thin near-wall layer. Retaining only the pressure gradient, the simplified wall-normal momentum equation is expressed as
\begin{equation}\label{wm-stress}
\frac{\partial}{\partial n} \left[ \left( \nu + \nu_t \right)
\frac{\partial u_i}{\partial n} \right]
= \frac{1}{\rho} \frac{\partial p}{\partial \tau_i}, \quad i=1,\,2,
\end{equation}
where $n$ is the wall-normal coordinate, $\tau_i$ ($i=1,\,2$) are the two orthogonal wall-tangential directions, and $u_i$ are the corresponding tangential velocity components to be solved on the embedded mesh. The density $\rho$ and the wall-parallel pressure gradient $\partial p/\partial \tau_i$ are treated as known forcing terms, provided directly by the outer solution at the exchange location. Furthermore, $\nu$ is the molecular kinematic viscosity, while the eddy viscosity $\nu_t$ requires an additional turbulence closure, specified below.

To close Eq.~\eqref{wm-stress}, the eddy viscosity $\nu_t$ is evaluated using the modified mixing-length model of \cite{WM-NEQ-2}, which builds on the pressure-gradient-based velocity scale introduced by \cite{WM-NEQ-3}. The eddy viscosity is expressed as
\begin{equation}\label{nu_t-NEQC}
    \nu_t = (\kappa y_n)^2|S|D_p(y_n),
\end{equation}
where $\kappa=0.41$ and $y_n$ is the wall-normal distance. Here $|S| = \sqrt{(\partial u_{1}/\partial n)^2+(\partial u_{2}/\partial n)^2}$ denotes the magnitude of the wall-normal shear rate. The term $D_p(y_n)$ is a modified damping function that incorporates the influence of the pressure gradient. The friction velocity $u_\tau$ and the pressure-gradient-based velocity $u_p$ are defined as
\begin{equation*}
u_\tau = \sqrt{\frac{|\tau_w|}{\rho}},\quad
u_p = \left( \frac{\nu}{\rho} \sqrt{ \left(\frac{\partial p}{\partial \tau_1}\right)^2
+ \left(\frac{\partial p}{\partial \tau_2}\right)^2 } \right)^{1/3},
\end{equation*}
where $u_p$ extends the one-dimensional form of \cite{WM-NEQ-2} to the two wall-tangential directions. A parameter $\alpha$ measures their relative dominance,
\begin{equation*}
    \alpha = \frac{u_\tau^2}{u_\tau^2+u_p^2},\quad
    \alpha_*=\max\left(\alpha,\ 0.5\left(\tanh(10(\alpha-0.5))+1\right)\right),
\end{equation*}
which yields a mixed friction velocity
$u_{\tau p}=\sqrt{u_\tau^2+(1-\alpha_*)u_p^2}$. The damping function is then
\begin{equation*}
    D_p(y_n) = (1-e^{-y_p^+/A_p^+})^2,
\end{equation*}
with the modified non-dimensional wall distance and effective damping coefficient
\begin{equation*}
    y_p^+=y_n u_{\tau p}/\nu,\quad A_p^+=1+\alpha_*^3(A^+-1),\quad A^+=26.
\end{equation*}
This formulation generalizes the standard van Driest damping function,
\begin{equation*}
    D(y_n)=\left(1-e^{-\frac{y^+}{A^+}}\right)^2,
    \qquad
    y^+=\frac{y_n u_\tau}{\nu},
\end{equation*}
by replacing the friction velocity $u_\tau$ in the wall-distance scaling with the mixed velocity $u_{\tau p}$, and the constant $A^+$ with the pressure-gradient-dependent effective coefficient $A_p^+$.

In this work, the wall model is used to reconstruct the wall shear stress and to provide the corresponding momentum flux to the outer compressible CGKS-5th solver. The density, molecular viscosity, and pressure-gradient term appearing in Eq.~\eqref{wm-stress} are evaluated from the CGKS-5th solution, while the eddy viscosity $\nu_t$ is determined by the closure model above.
This approximation is acceptable for the subsonic and transonic separated flows studied in this work, where the pressure gradient is the dominant non-equilibrium effect. Stronger compressibility brings in effects beyond the present formulation, including significant convective and unsteady terms, a non-negligible wall-normal pressure gradient, and heat transfer. Capturing these calls for a more complete wall model, which we leave to future work.

\subsection{Analysis of pressure-gradient corrections and the choice of exchange location}
\label{subsec:regimes} 

A classical equilibrium wall model neglects the wall-parallel pressure gradient, setting the right-hand side of Eq.~\eqref{wm-stress} to zero. This omission is the root of its shortcomings under strong pressure gradients: it cannot let the wall shear stress fall toward zero at separation, nor reproduce the curvature that the pressure gradient imparts to the near-wall velocity profile, leaving an inaccurate wall-stress estimate. A further limitation arises from the van Driest damping function, whose $u_\tau$ scaling causes it to collapse as $\tau_w\to0$ near separation and to over-suppress the modeled eddy viscosity.
The model formulated in subsection~\ref{subsec:formulation} accounts for the pressure gradient through the source term in Eq.~\eqref{wm-stress} and the modified damping function. How these act depends on the near-wall state, defining the two regimes analyzed below.

The first regime is a turbulent boundary layer under an adverse pressure gradient, where the eddy viscosity is significant and the wall model acts as a turbulence closure for the unresolved inner layer. By retaining the pressure-gradient source term, Eq.~\eqref{wm-stress} enables the modeled wall shear stress to respond to the adverse gradient and, when supported by the exchange-location data, to approach zero or even change sign, allowing the boundary condition to represent incipient separation and near-wall backflow; under a vanishing pressure gradient, the source term drops out and the equilibrium balance is recovered. In the same near-separation limit, where $\tau_w\to0$, the modified damping function prevents the standard van Driest form from collapsing and driving the modeled eddy viscosity to zero.

The second regime occurs in a laminar near-wall flow under a pressure gradient. Within the present framework, the mixing-length closure naturally yields a negligible eddy viscosity in this laminar region, and Eq.~\eqref{wm-stress} reduces to a balance between wall-normal viscous diffusion and the pressure-gradient source,
\begin{equation*}\label{wm-laminar}
\nu\,\frac{\partial^2 u_i}{\partial n^2} = \frac{1}{\rho}\frac{\partial p}{\partial \tau_i}.
\end{equation*}
The near-wall velocity profile is therefore curved, with a curvature proportional to the pressure gradient. Under a favorable gradient ($\partial p/\partial \tau_i<0$), as in the accelerated laminar flow ahead of separation, this curvature is negative. A no-model estimate, which assumes a uniform velocity variation up to the exchange location, then underestimates the wall-normal gradient at the wall, and hence the wall shear stress, with an error that grows on coarser meshes. By retaining the pressure-gradient source term, the wall model reconstructs the curved profile and recovers the correct wall gradient, acting here not as a turbulence closure but as a correction to the near-wall velocity profile.

In the present non-equilibrium wall-model formulation, the tangential pressure gradient is assumed to be invariant in the wall-normal direction across the wall-modeled layer. The exchange location is therefore selected according to the local flow configuration so that this approximation remains consistent with the thickness and pressure variation of the modeled layer.

For flows in which the turbulent boundary layer remains thin and largely attached, such as the transonic high-Reynolds-number flow over the RAE 2822 airfoil in this work, where separation in the shock region is limited to a small local bubble, the wall-normal variation of the tangential pressure gradient across the modeled layer is expected to be small. The exchange location is placed at the third off-wall cell, kept thin enough for the wall-normal-invariant pressure-gradient assumption to hold, and the model operates in the first regime described above.

For flows with laminar separation, such as the subsonic flow over a circular cylinder in this work, the boundary layer remains laminar up to separation, beyond which it detaches into a thick recirculating wake with significant wall-normal pressure variation. Modeling a thick wall-normal layer there would make the assumption of a wall-normal-invariant tangential pressure gradient questionable. The exchange location is therefore set at the first off-wall cell over the whole body, so that the modeled layer remains thin and the pressure-gradient approximation is applied only over a short wall-normal distance. Although matching below the logarithmic layer introduces additional sensitivity into the wall-stress prediction, it avoids the potentially larger modeling error of applying the wall-normal-invariant assumption across a thick recirculating layer. Near the leading edge ahead of separation, the laminar boundary layer accelerates under a favorable pressure gradient and falls into the second regime described above.

\subsection{Solution of the wall model and coupling to CGKS-5th}

Equation~\eqref{wm-stress} is nonlinear, since its diffusion coefficient $\nu+\nu_t$ depends on the unknown velocity field through the shear rate $|S|$ and the damping function $D_p$, and is therefore solved iteratively. The tangential velocity components $u_i$ are discretized on the one-dimensional embedded mesh using a second-order cell-centered finite volume method, with Dirichlet conditions imposed at both the wall and the exchange location. At each nonlinear iteration the eddy viscosity is updated from the previous iterate, giving a tridiagonal system that is solved with the Thomas algorithm, and the iteration continues until the $L_\infty$ norm of the velocity update falls below a prescribed tolerance. Upon convergence, the wall shear-stress components are evaluated from the wall-normal velocity gradient,
\begin{equation}
    \tau_{w,i}
    =
    \mu
    \left.
    \frac{\partial u_{i}}{\partial n}
    \right|_{\mathrm{wall}},
    \qquad i=1,2.
\end{equation}
With the wall model, this gradient is taken from the near-wall profile reconstructed on the embedded mesh; without the wall model, it is estimated directly from the first off-wall cell.

The modeled wall shear stress is coupled back to the outer solver through the wall momentum flux alone. The inviscid part of the wall momentum flux, together with all other fluxes, is provided by the CGKS-5th, which evaluates inviscid and viscous fluxes within a single formulation, so that all of them are produced together in one flux evaluation. The wall momentum flux is then assembled by adding the modeled viscous stress $\tau_{w,i}$ to the inviscid part. The overall workflow of the proposed wall-modeled CGKS framework is outlined in Algorithm~\ref{WM-CGKS5-algorithm}.

\begin{algorithm}[!h]
\SetAlgoLined
Initialize flow field\;
Compute reconstruction polynomial coefficients on a standard cell\;
Select exchange location and embed a 1D wall-normal grid for each wall interface\;
Initialize the wall-model velocity profile on each embedded grid\;
\While{\rm{TIME} $ \leq $ \rm{TSTOP}}{
    Compute time step $\Delta t$\;
    \textcolor{customgreen}{\textit{! \textbf{Stage 1}}}\\
    Apply boundary condition at $t=t^n$\;
    \For{each cell interface}
    {
    \For{each side of the interface}
    {
    Transform variables and gradients to the reference coordinate system\;
    Assemble polynomials $P^4(\boldsymbol{\xi})$ and $P_m^1(\boldsymbol{\xi})$\;
    Apply GENO for nonlinear combination\;
    Transform reconstructed values back to the physical coordinate system\;
    }
    Compute flux and pointwise conservative variables using GKS\;
    \If{the interface lies on the wall}{
    Output the inviscid flux using GKS\;}
    }
    \For{each wall interface}
    {
    Extract flow variables and gradients at the exchange location\;
    Project them onto the tangential directions\;
    Solve wall model Eq.~\eqref{wm-stress} and obtain the wall shear stress\;
    Project the wall shear stress back to global coordinates\;
    Assemble the wall momentum flux from the inviscid flux and the modeled wall shear stress\;
    }
    Update $\boldsymbol{Q}(t^{*})$, $(\nabla \boldsymbol{Q})(t^{*})$, and $(\partial_{l} \boldsymbol{Q})(t^{*})$\;
    \textcolor{customgreen}{\textit{! \textbf{Stage 2: similar to Stage 1}}}\\
    Apply boundary condition at $t=t^*$\;
    Perform reconstruction and flux calculation\;
    Solve wall model and assemble the wall momentum flux\;
    Update $\boldsymbol{Q}(t^{n+1})$, $(\nabla \boldsymbol{Q})(t^{n+1})$, and $(\partial_{l} \boldsymbol{Q})(t^{n+1})$\;
}
\caption{\label{WM-CGKS5-algorithm} Computational workflow of the wall-modeled CGKS-5th}
\end{algorithm}

\section{Numerical verification}

To validate the proposed wall-modeled CGKS framework, this section considers two representative three-dimensional turbulent flows: subsonic low-Reynolds-number flow past a circular cylinder and transonic high-Reynolds-number flow over an RAE 2822 airfoil. Spanning several orders of magnitude in Reynolds number and involving distinct separation mechanisms, these two cases enable a systematic assessment of the framework's accuracy and robustness across different flow regimes.

In this study, the embedded one-dimensional wall-normal grid consists of 32 cells, and the wall-model input variables are taken from the cell-averaged values at the exchange location. The nonlinear iteration is initialized differently for the two cases. For the RAE 2822 case, the converged profile from the previous time step is reused. For the cylinder case, the strong temporal variation in the separated region makes this reuse less reliable, and the embedded profile is instead re-initialized at each time step from a linear distribution based on the current exchange-location velocity, which is found to improve robustness. For convenience, the wall-modeled fifth-order compact gas-kinetic scheme is denoted WM CGKS-5th.

For the evolution of the flow fields, the time step $\Delta t$ is given by the CFL condition. In all test cases, the CFL number is taken as $\text{CFL} > 0.5$. 
For viscous flows, the time step is also limited by the viscous term as $\Delta t = \text{CFL}\cdot h^2/(3\nu)$, where $h$ is the cell size and $\nu$
is the kinematic viscosity coefficient. 
The collision time is related to the viscosity coefficient,
\begin{align*}
\tau=\frac{\mu}{p}+C \displaystyle|\frac{p_l-p_r}{p_l+p_r}|\Delta t,
\end{align*}
where $\mu$ is the dynamic viscous coefficient, $p$ is the pressure at the cell interface, $p_l$ and $p_r$ denote the pressure on the left and right sides of the cell interface, and $C=10.0$. In smooth flow regions, the collision time reduces to
\begin{equation*}
\tau=\frac{\mu}{p}.
\end{equation*}
In this section, the specific heat ratio $\gamma$ takes 1.4.

All numerical simulations in this section are performed using a multi-GPU implementation of the CGKS-5th and WM CGKS-5th schemes. The solvers are parallelized via CUDA for on-device computation and MPI for inter-process communication. This framework facilitates large-scale simulations, deploying eight NVIDIA GeForce RTX 4090 GPUs with double-precision arithmetic.

\begin{figure}[!h]
\centering
\includegraphics[width=0.465\textwidth]{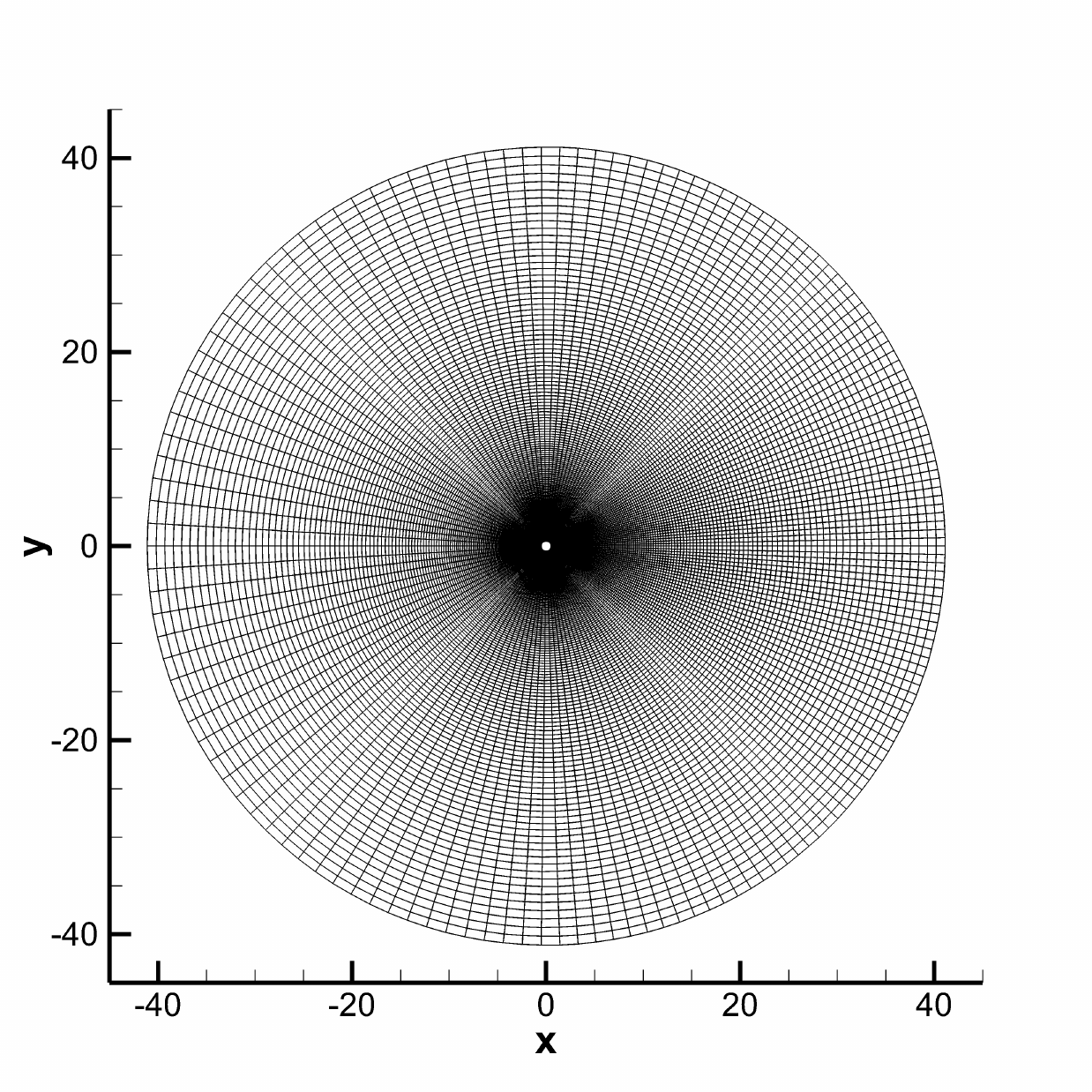}
\includegraphics[width=0.465\textwidth]{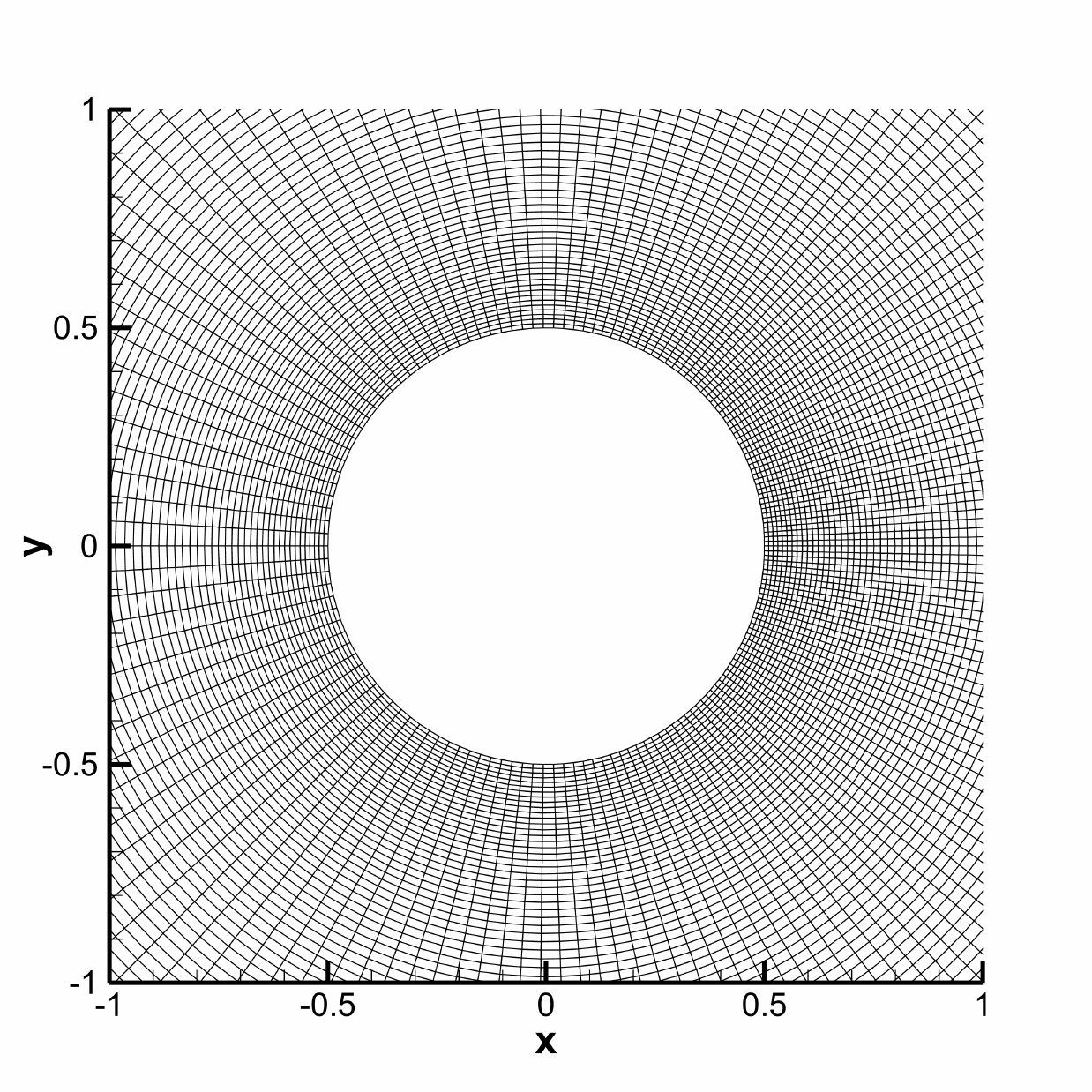}
\caption{\label{cylinder-mesh} Turbulent flow past a cylinder: visualization of the computational mesh in the x-y plane with a full view (left) and a detailed view (right).}
\end{figure}

\begin{figure}[!h]
\centering
\includegraphics[width=0.49\textwidth]{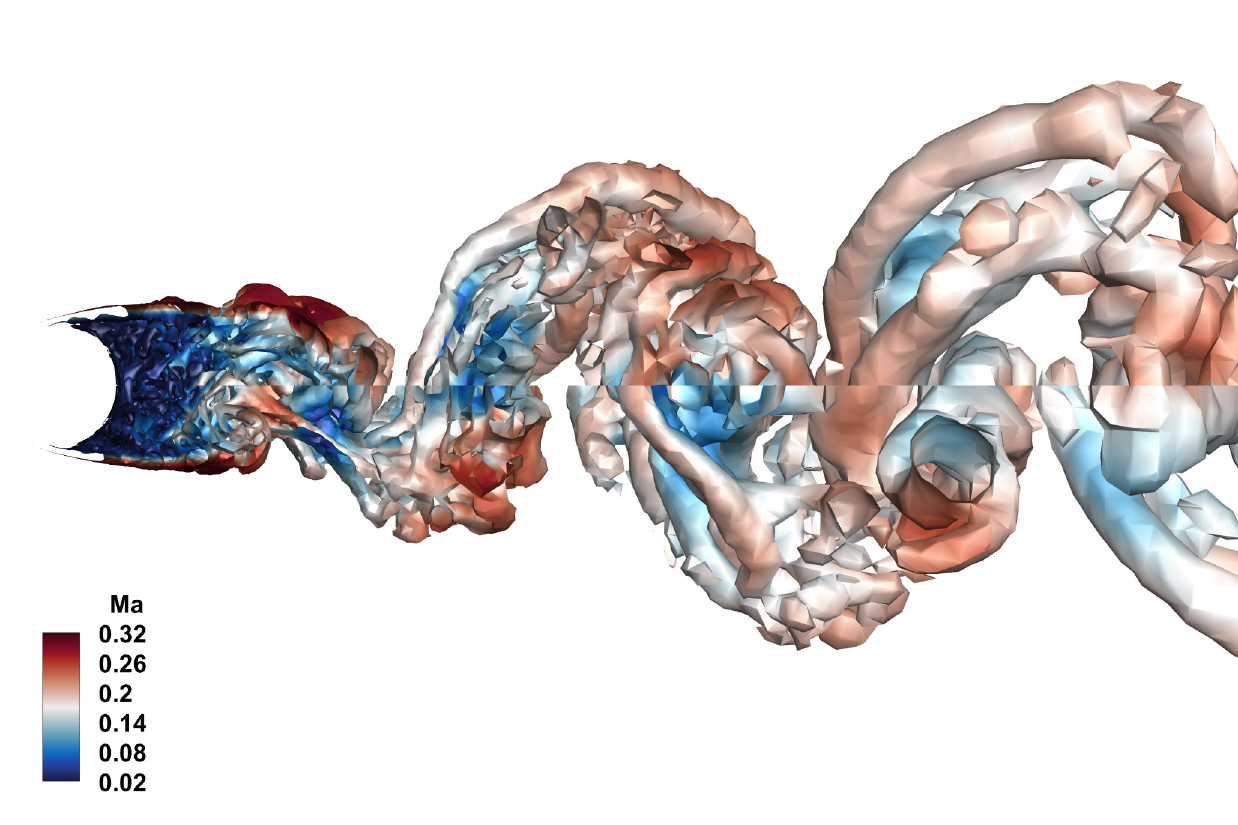}
\includegraphics[width=0.49\textwidth]{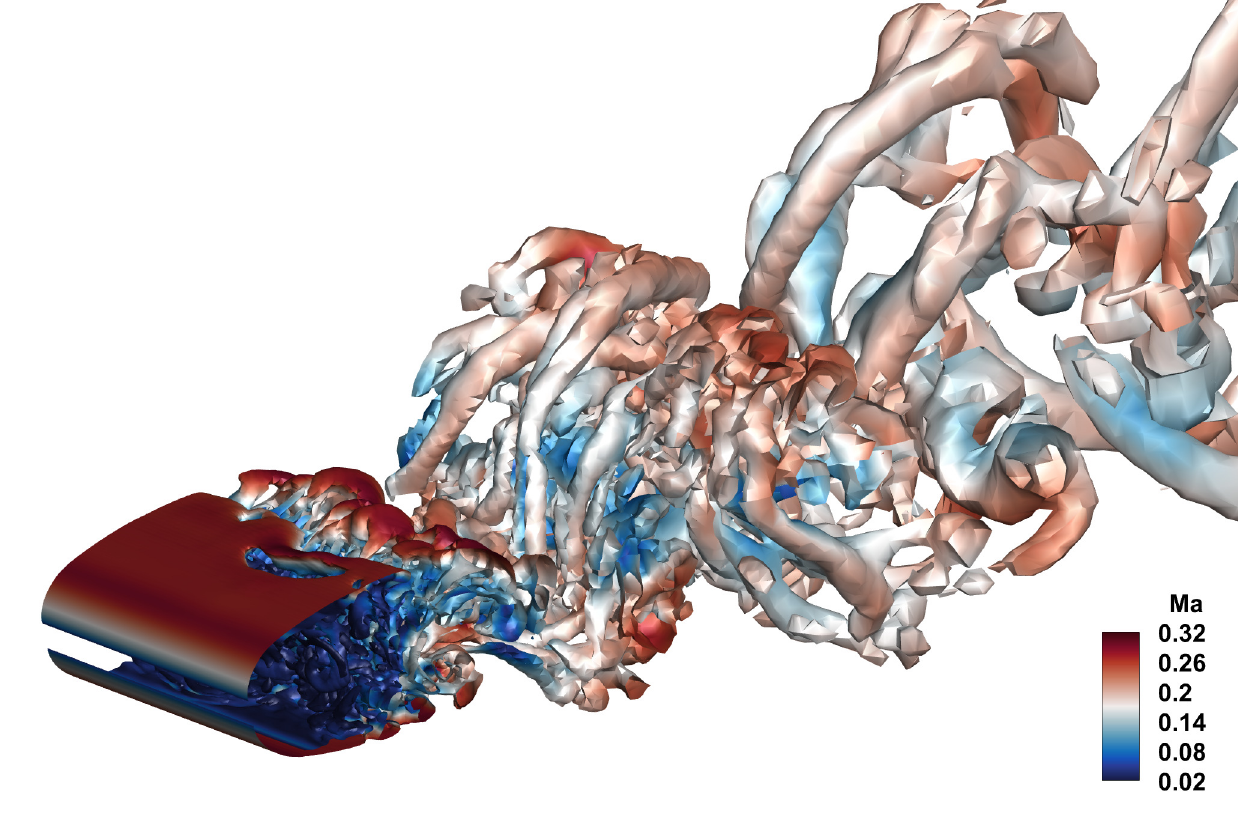}
\caption{\label{cylinder-vortex} Turbulent flow past a cylinder: the iso-surface of the Q-criterion ($Q=0.01$) with a front view (left) and an isometric view (right).}
\end{figure}

\begin{figure}[!h]
\centering
\includegraphics[width=0.495\textwidth]{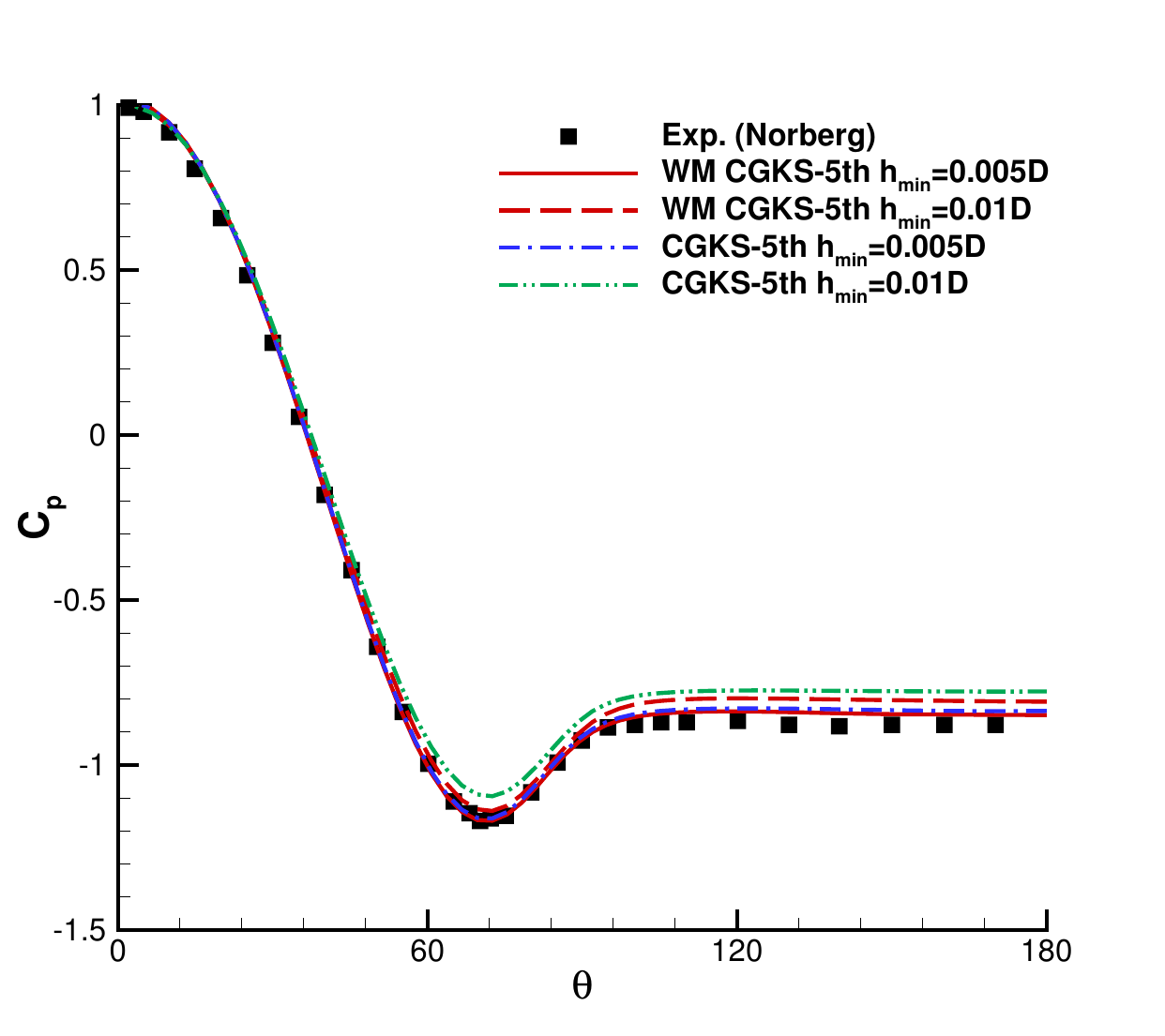}
\includegraphics[width=0.495\textwidth]{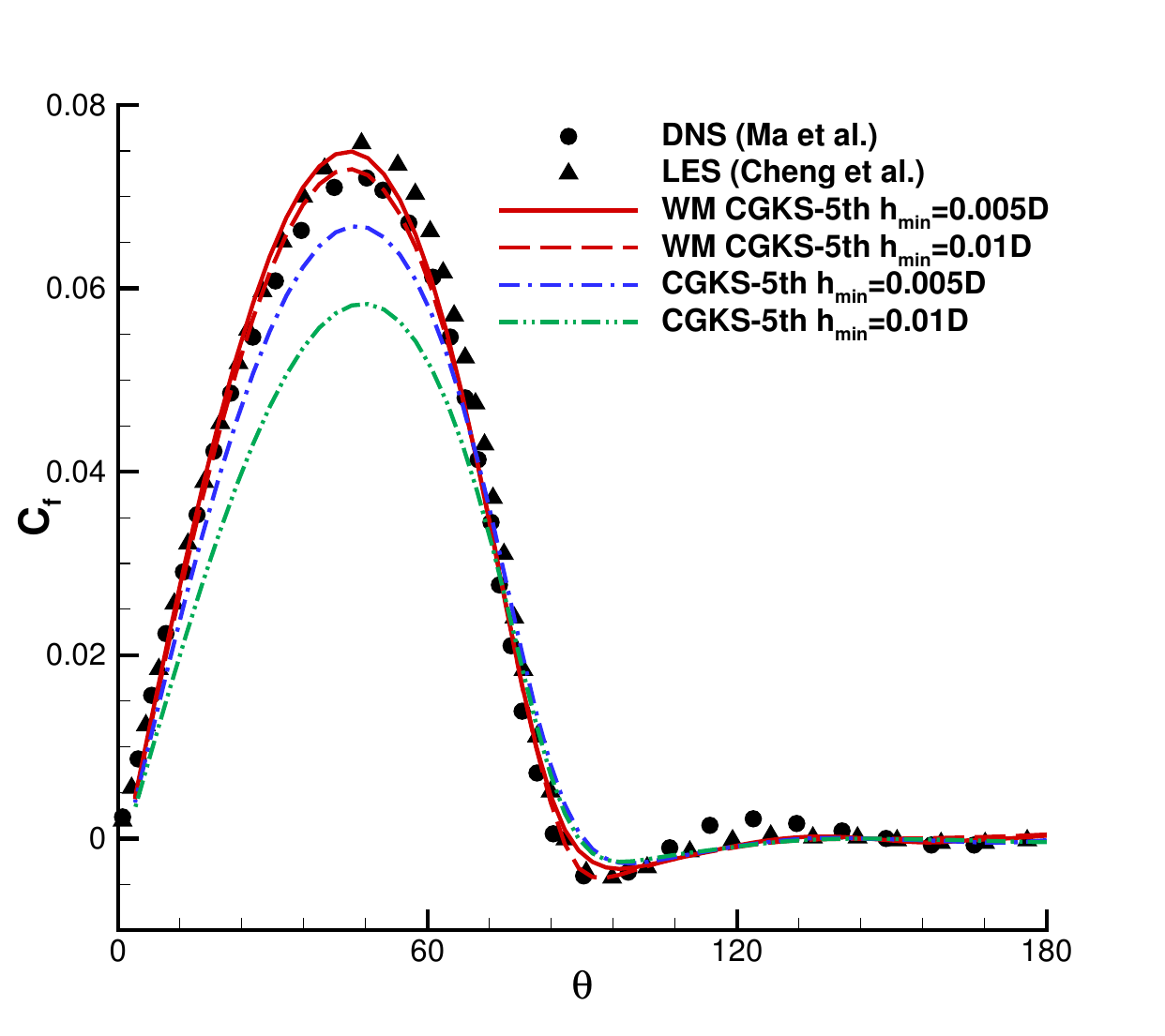}
\caption{\label{cylinder-cp-cf} Turbulent flow past a cylinder: the pressure coefficient (left) and skin-friction coefficient (right) along the cylinder surface obtained by CGKS-5th and WM CGKS-5th using linear reconstruction.}
\end{figure}

\begin{figure}[!h]
\centering
\includegraphics[width=0.72\textwidth]{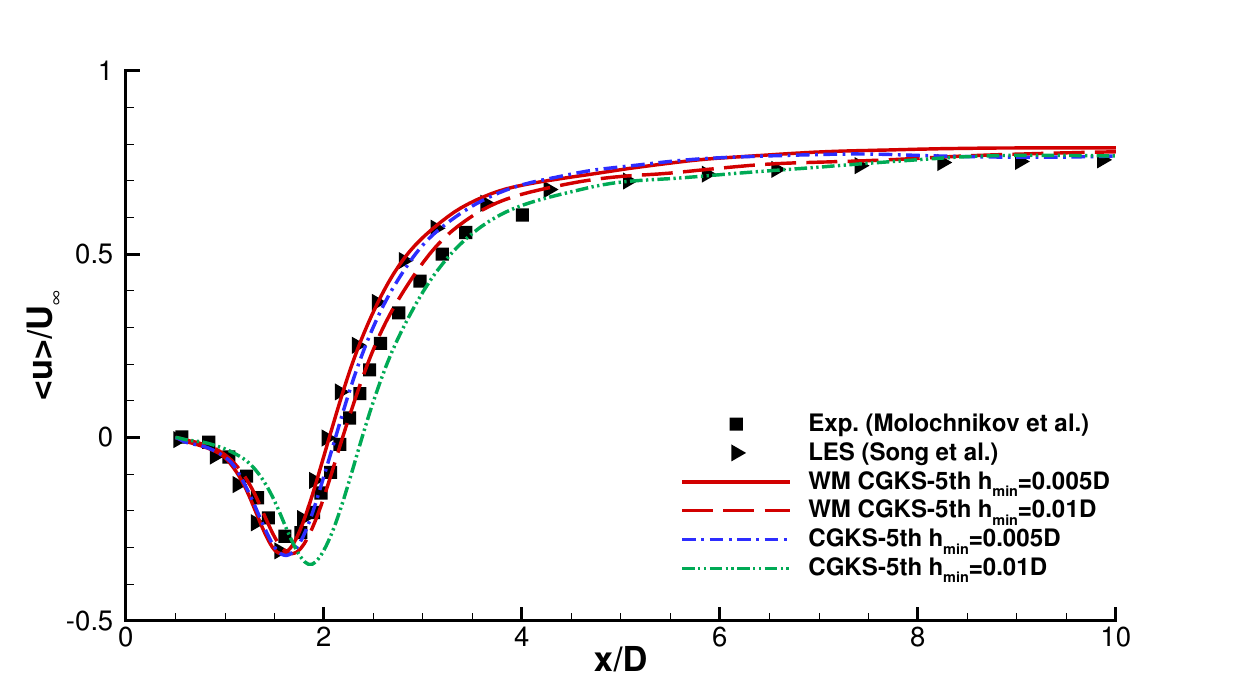}
\caption{\label{cylinder-y0} Turbulent flow past a cylinder: the streamwise velocity profiles along the centerline in the cylinder wake obtained by CGKS-5th and WM CGKS-5th using linear reconstruction.}
\end{figure}

\begin{figure}[!h]
\centering
\includegraphics[width=0.496\textwidth]{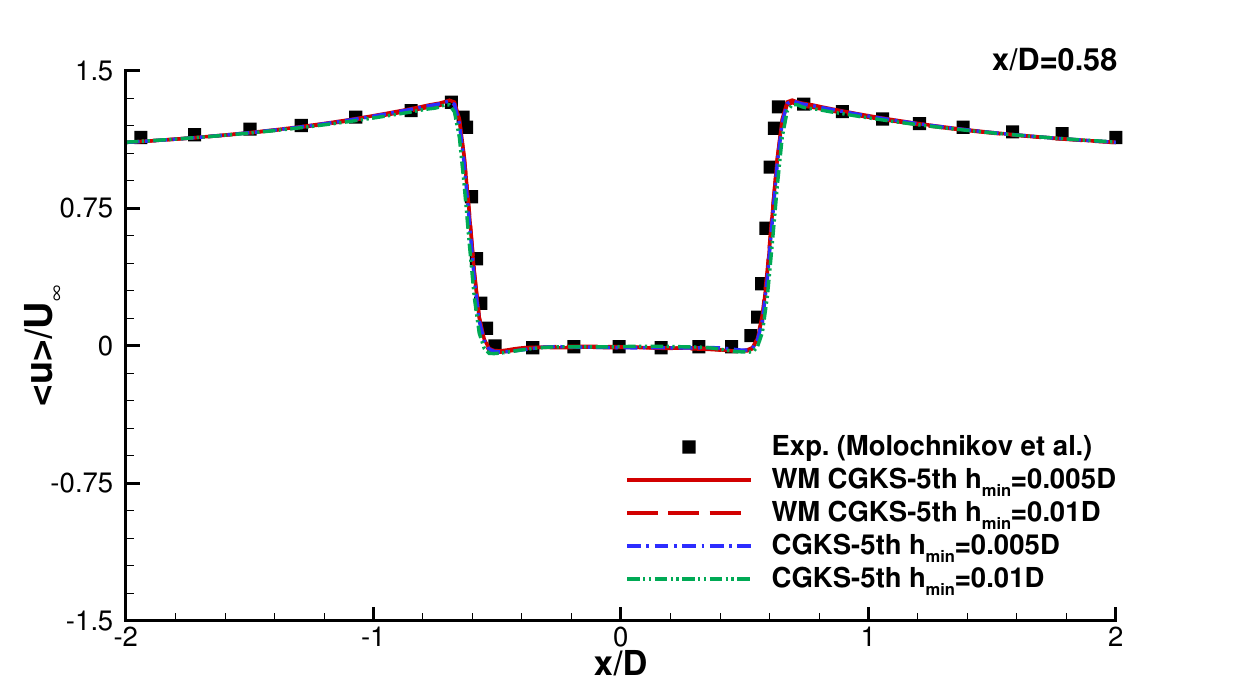}
\includegraphics[width=0.496\textwidth]{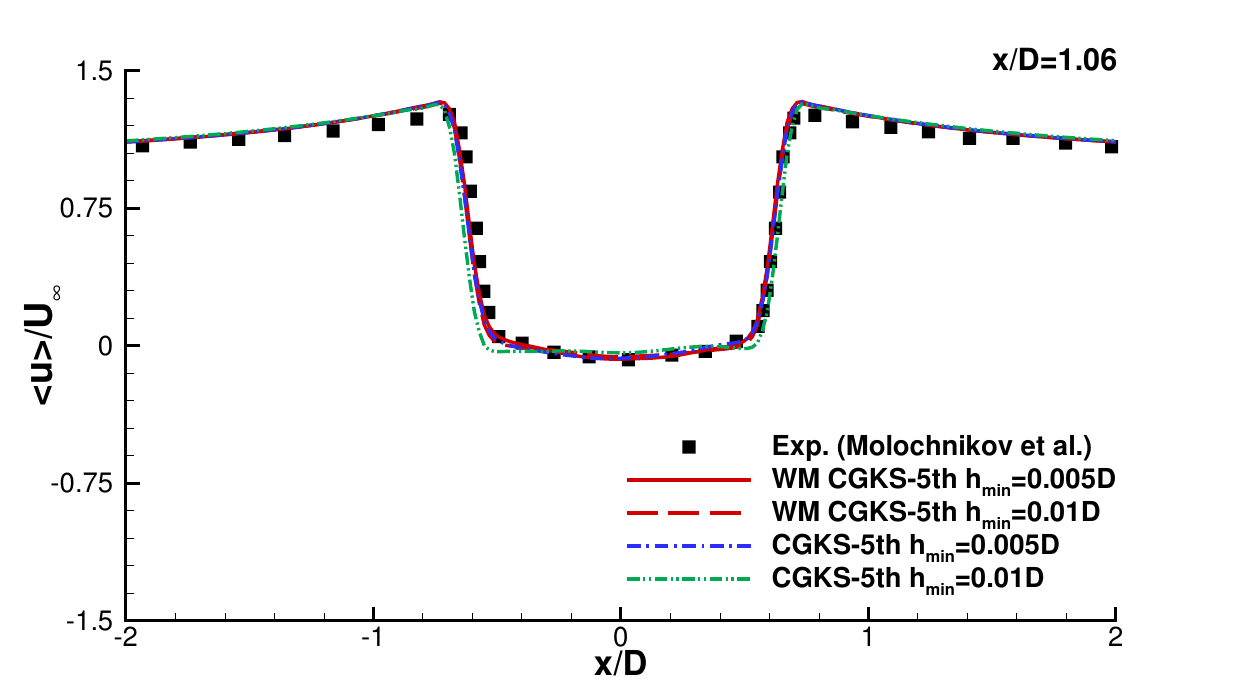}\\
\includegraphics[width=0.496\textwidth]{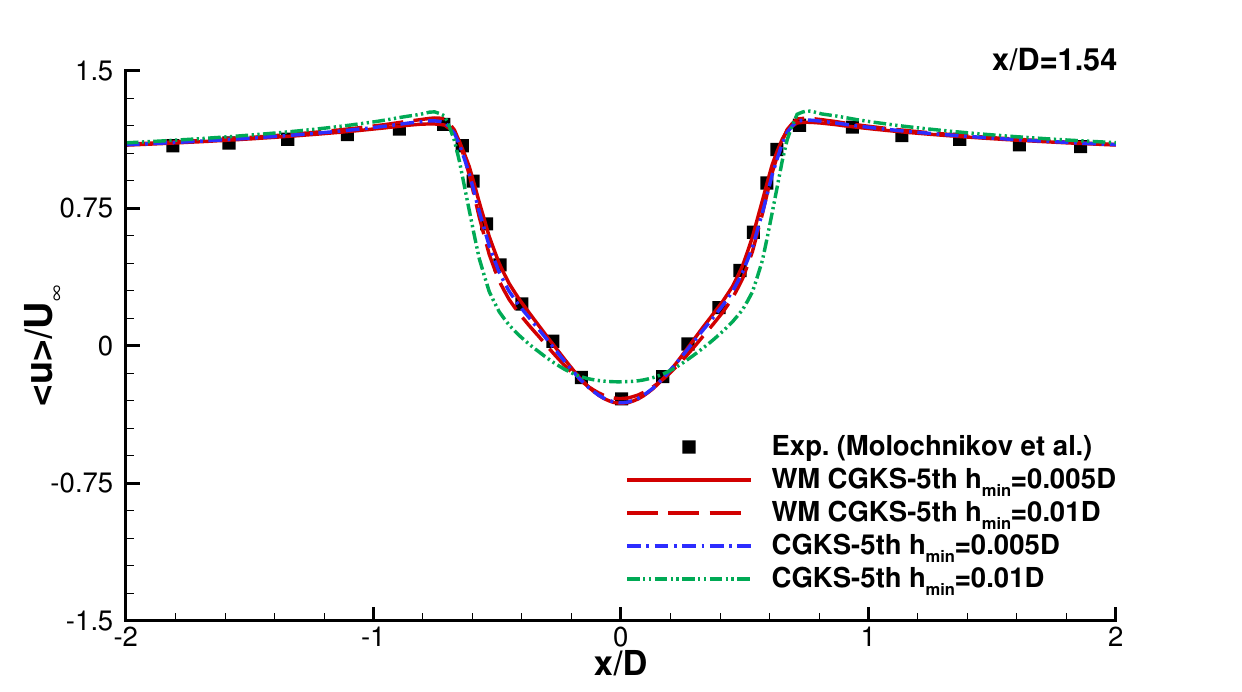}
\includegraphics[width=0.496\textwidth]{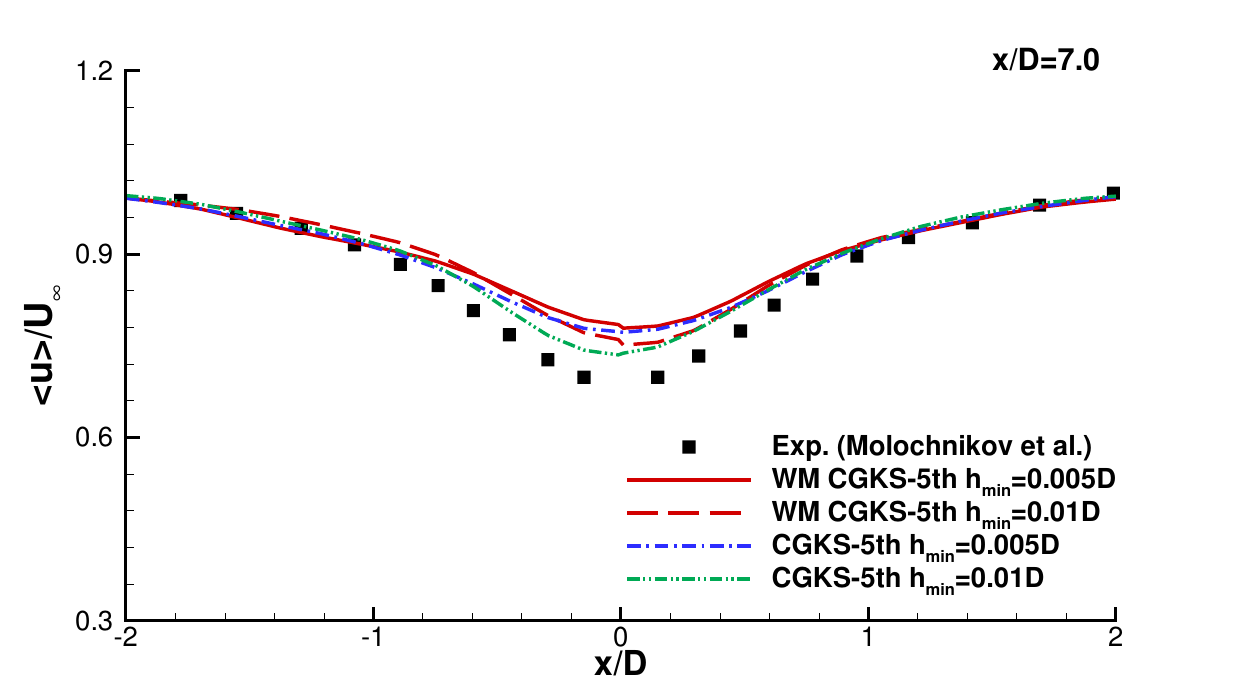}
\caption{\label{cylinder-o1} Turbulent flow past a cylinder: the streamwise velocity profiles at different cylinder wake locations obtained by CGKS-5th and WM CGKS-5th using linear reconstruction.}
\end{figure}

\begin{figure}[!h]
\centering
\includegraphics[width=0.496\textwidth]{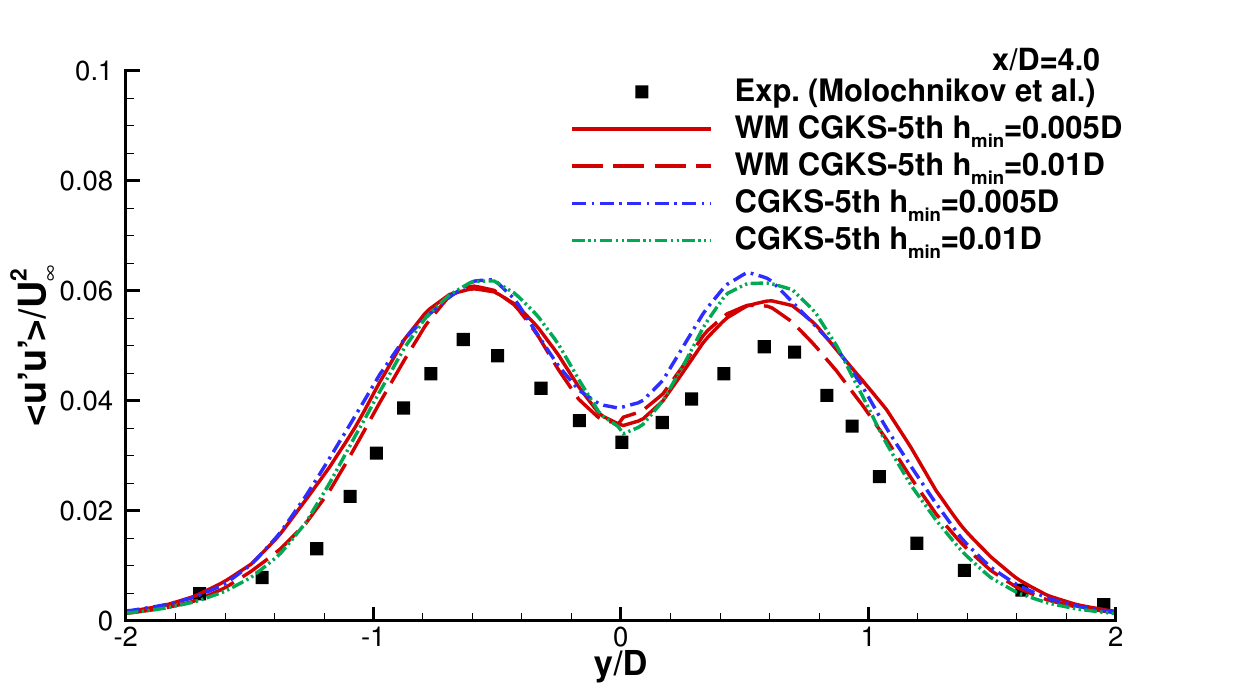}
\includegraphics[width=0.496\textwidth]{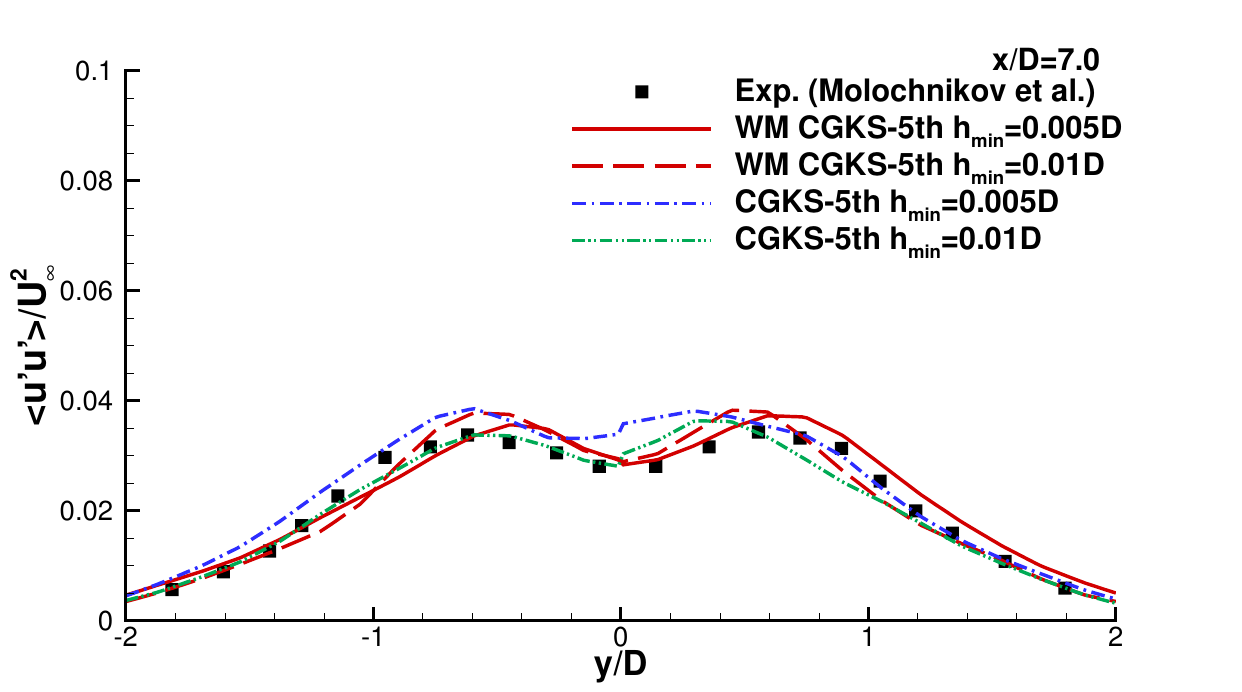}\\
\includegraphics[width=0.496\textwidth]{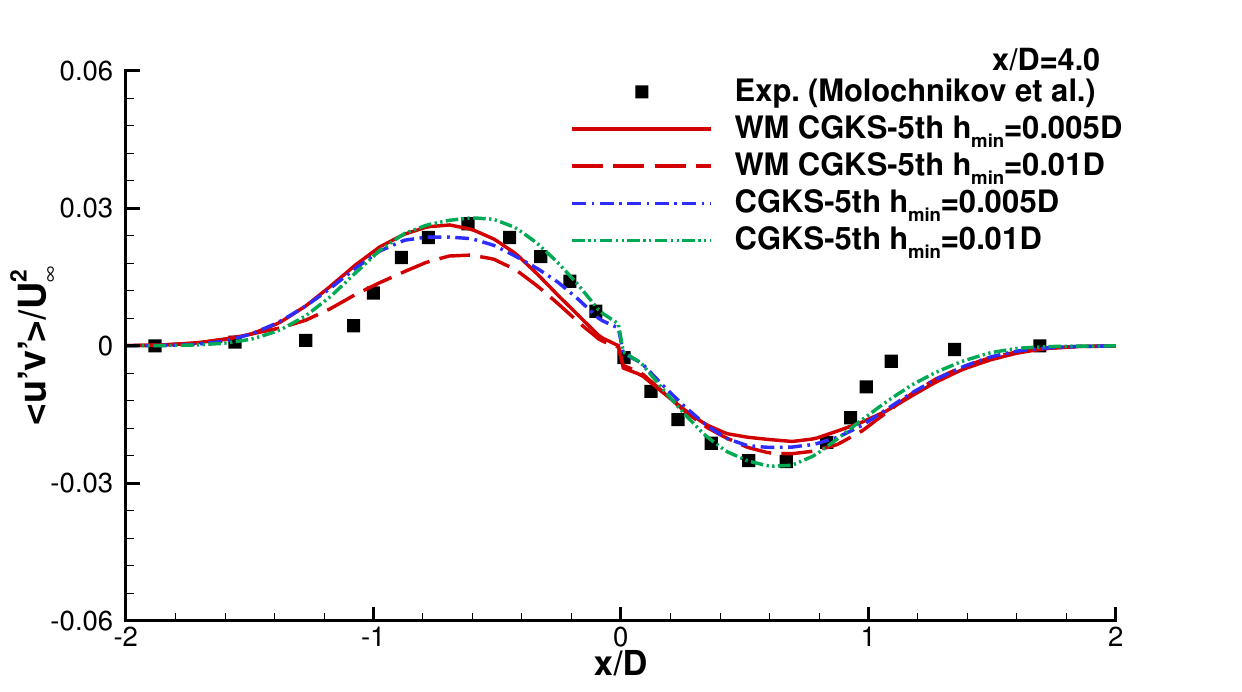}
\includegraphics[width=0.496\textwidth]{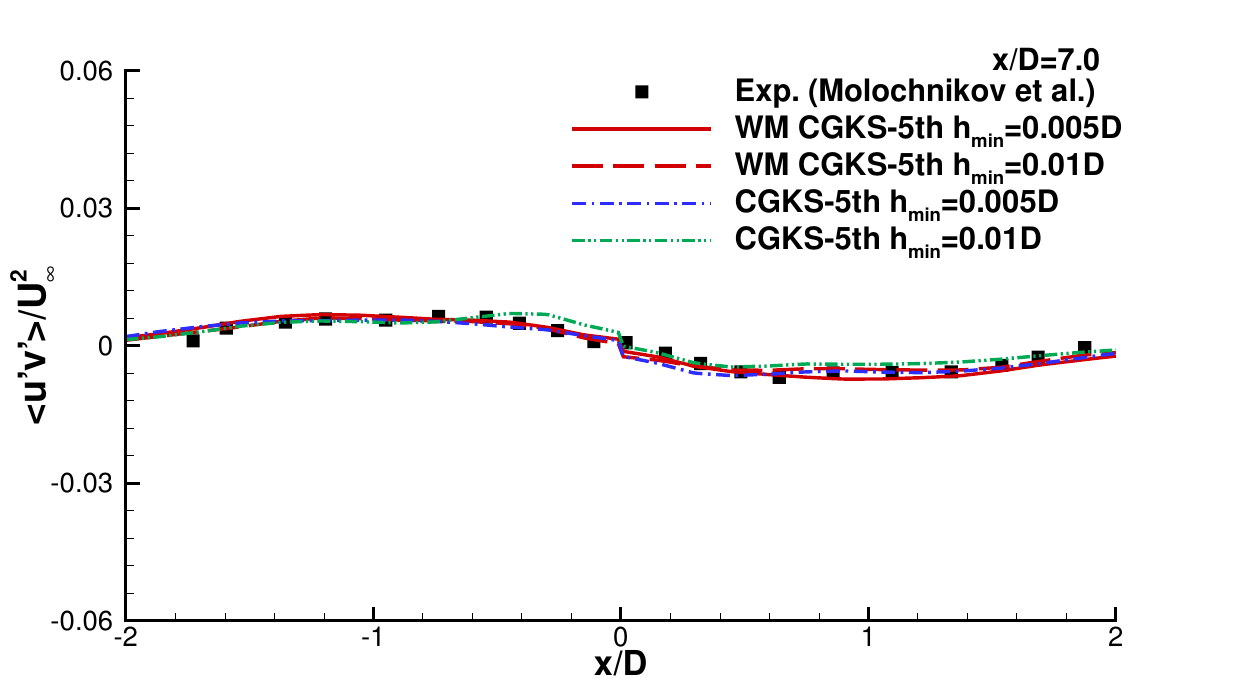}\\
\includegraphics[width=0.496\textwidth]{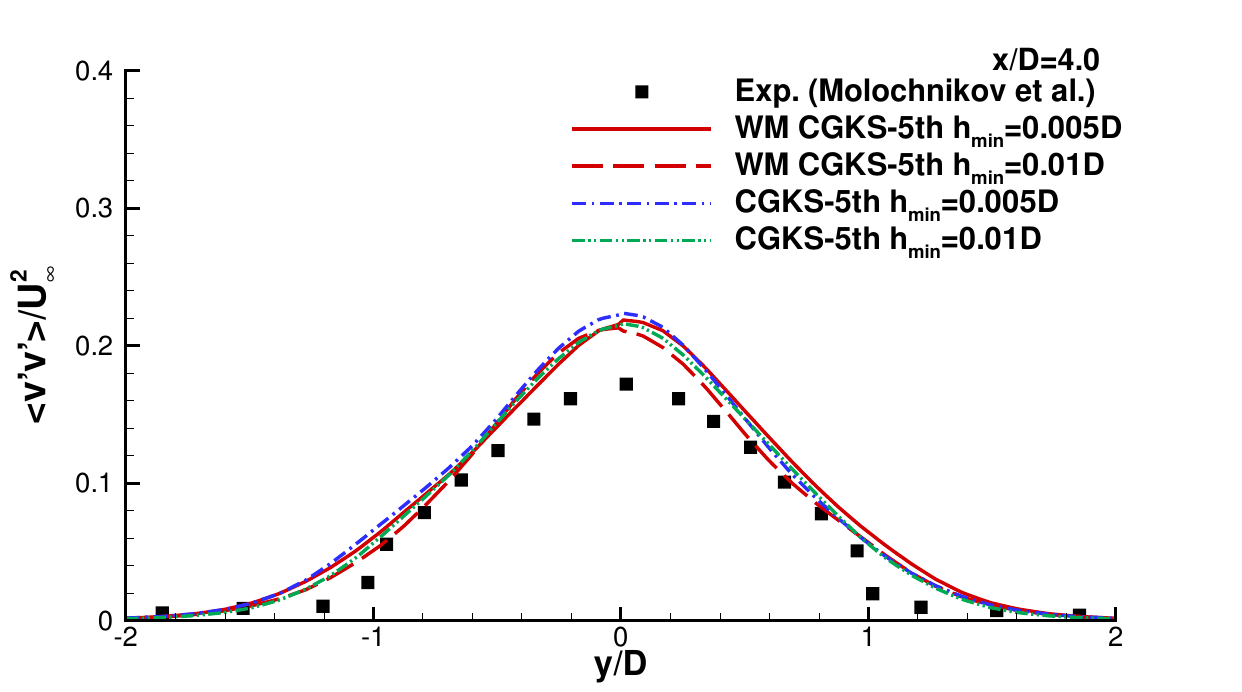}
\includegraphics[width=0.496\textwidth]{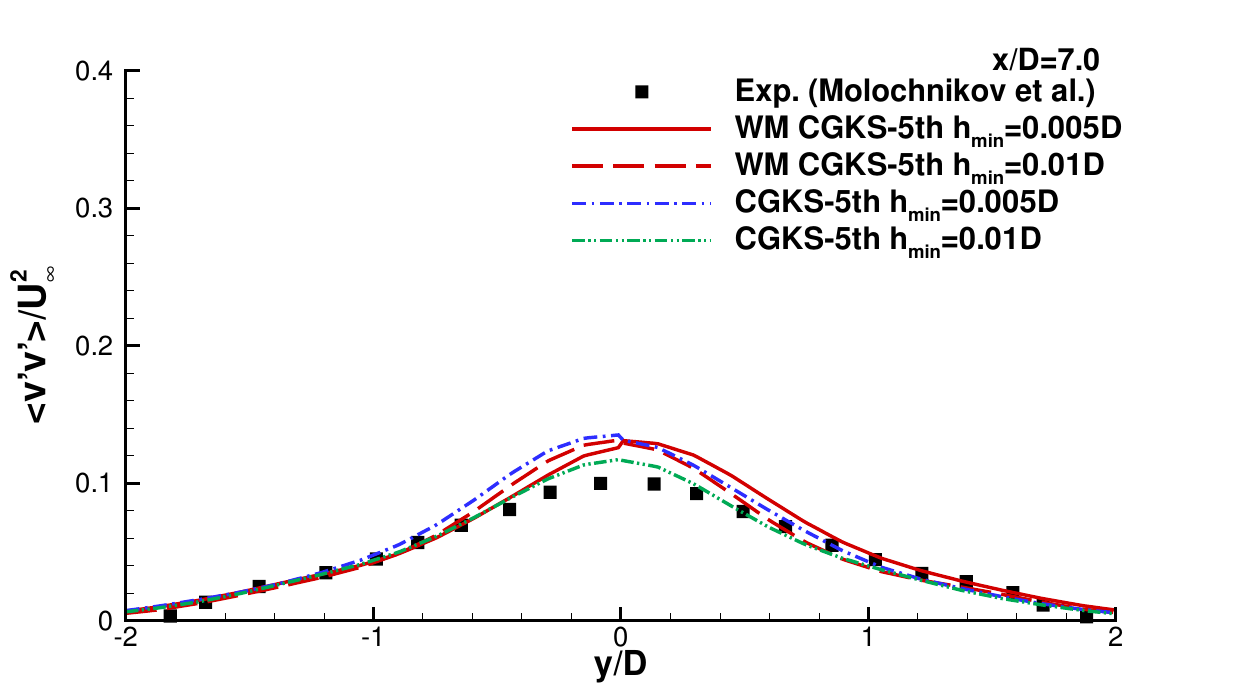}
\caption{\label{cylinder-o2} Turbulent flow past a cylinder: the Reynolds stress components profiles at different cylinder wake locations obtained by CGKS-5th and WM CGKS-5th using linear reconstruction.}
\end{figure}

\subsection{Turbulent flow past a cylinder}

A subsonic flow past a circular cylinder at $Re_D=3900$ is first considered to assess the performance of the WM CGKS-5th. At this subcritical Reynolds number, the boundary layer remains laminar up to separation and the detached shear layers transition to turbulence in the wake; the configuration thus combines laminar separation with a turbulent wake, providing a bluff-body benchmark for the wall-modeled framework.

The cylinder, with a diameter of $D=1$, is centered at the origin.
The three-dimensional computational domain is discretized using an O-type structured mesh.
The outer boundary is concentric with the cylinder at a radial distance of approximately $40D$, and the spanwise extent is $L_z=\pi D$ with a periodic boundary condition.
The free-stream Mach number is $Ma_\infty=0.2$ and the Reynolds number is $Re_D=3900$, based on the cylinder diameter. The flow field is initialized with the uniform free-stream condition
\begin{align*}
(\rho,U,V,W,p)_{\infty} = (1, Ma_\infty,0,0, 1/\gamma).
\end{align*}
A no-slip adiabatic wall boundary condition is imposed on the cylinder surface, and a far-field condition is applied at the outer boundary.

To evaluate both the coarse-grid performance and the grid convergence, two meshes are employed.
The coarse mesh consists of $200\times160\times60$ cells in the radial, circumferential, and spanwise directions, respectively, with a minimum wall-normal spacing of $h_{\min}=0.01D$.
For the grid convergence study, a refined mesh is generated by increasing the radial resolution, yielding $224\times160\times60$ cells and a reduced first-layer spacing of $h_{\min}=0.005D$.
In both configurations, the grid points are clustered in the near-wake region to adequately resolve the vortex shedding and are gradually stretched toward the far-field boundary, while the spanwise spacing remains uniform.
A cross-sectional view of the coarse mesh in the $x$-$y$ plane is shown in Figure~\ref{cylinder-mesh}. Both the CGKS-5th and the WM CGKS-5th employ linear reconstruction in this subsection.

Figure~\ref{cylinder-vortex} shows the instantaneous vortex structures generated by the WM CGKS-5th on the coarse mesh ($h_{\min}=0.01D$), visualized via Q-criterion isosurfaces ($Q=0.01$) colored by the Mach number.
The intricate small-scale vortices in the wake are clearly resolved, demonstrating the strong resolving capability of the scheme on a relatively coarse mesh.

Figure~\ref{cylinder-cp-cf} compares the surface distributions of the pressure coefficient $C_p$ and the skin-friction coefficient $C_f$ predicted by the CGKS-5th and the WM CGKS-5th against experimental~\cite{Cylinder-ref-cp}, DNS~\cite{Cylinder-ref-cf-dns}, and LES~\cite{Cylinder-ref-cf-les} reference data. For the $C_p$ distribution, both schemes reproduce the overall pressure trend around the cylinder, while the WM CGKS-5th yields closer agreement with the experimental data, particularly in the suction-peak magnitude and the base pressure.
A similar improvement is observed in the $C_f$ distribution.
Both schemes capture the general variation of the wall shear stress, but the WM CGKS-5th predicts the peak near $\theta\approx 45^\circ$ markedly more accurately.
On the coarse mesh with $h_{\min}=0.01D$, the WM CGKS-5th predicts $C_{f,\max}\approx 0.073$ in close agreement with the reference data, whereas the CGKS-5th underestimates it as $0.058$.
This peak lies in the strongly accelerated region upstream of separation and corresponds precisely to the regime discussed in Section~\ref{sec:wall_model}, in which retaining the pressure-gradient source term allows the wall model to recover the correct peak that a no-model linear estimate would underpredict. Furthermore, as the minimum wall-normal spacing $h_{\min}$ decreases, the WM CGKS-5th results progressively converge toward the reference solutions, confirming the grid convergence of the scheme.

Figure~\ref{cylinder-y0} shows the mean streamwise velocity profile $\langle u\rangle/U_\infty$ along the wake centerline, where the operator $\langle\cdot\rangle$ denotes temporal averaging combined with spatial averaging in the $z$-direction.
Results from the CGKS-5th and the WM CGKS-5th with linear reconstruction are compared with experimental data~\cite{Cylinder-ref-2} and LES results~\cite{Cylinder-ref-1} computed on a mesh of $512\times512\times128$ cells.
Both schemes correctly reproduce the reverse-flow region and the subsequent downstream recovery, while the WM CGKS-5th shows noticeably better quantitative agreement with the reference data, particularly in the recirculation length.

Figure~\ref{cylinder-o1} further presents the mean streamwise velocity profiles at four wake locations ($x/D=0.58$, $1.06$, $1.54$, and $7.0$), compared with experimental data~\cite{Cylinder-ref-2}.
At all stations, both schemes capture the overall wake-profile shape reasonably well, while the WM CGKS-5th provides more accurate predictions in the near-wake region.
Figure~\ref{cylinder-o2} presents the profiles of the Reynolds-stress components $\langle u'u'\rangle/U_\infty^2$, $\langle u'v'\rangle/U_\infty^2$, and $\langle v'v'\rangle/U_\infty^2$ at two wake locations ($x/D=4.0$ and $7.0$), compared with experimental data~\cite{Cylinder-ref-2}.
The predicted profiles agree well with the measurements at both stations, with only minor differences among the schemes and mesh resolutions, consistent with the far wake being governed by the outer solver and largely insensitive to the near-wall treatment.

Table~\ref{cylinder-cd-theta} quantitatively compares the mean drag coefficient $C_d$ and the separation angle $\theta_{\text{sep}}$ obtained by the CGKS-5th and the WM CGKS-5th against experimental measurements~\cite{Cylinder-ref-cd, Cylinder-ref-theta}.
The CGKS-5th overpredicts the separation angle, whereas the WM CGKS-5th brings it close to the measured value; on the refined mesh it attains the best agreement in both $C_d$ and $\theta_{\text{sep}}$ and shows excellent grid convergence. These results further validate the accuracy of the proposed wall-modeled framework in predicting boundary-layer separation.

\begin{table}[!h]
\begin{center}
\def\temptablewidth{0.9\textwidth}{\rule{\temptablewidth}{1.0pt}}
\begin{tabular*}{\temptablewidth}{@{\extracolsep{\fill}}c|c|c|c} 
Scheme & Mesh  & $C_d$ & $\theta_{\text{sep}}(\text{deg})$   \\
\hline
Exp.(Norberg) & - & 0.98 & -  \\
Exp.(Parnaudeau et al.) & - & - & 88  \\
WM CGKS-5th & $224\times 160\times 60~ (h_{min}=0.005D)$ & 0.974  & 87.7\\
WM CGKS-5th & $200\times 160\times 60~ (h_{min}=0.01D)$ & 0.932  & 86.2 \\ 
CGKS-5th & $224\times 160\times 60~ (h_{min}=0.005D)$ & 0.965  & 90.2 \\
CGKS-5th & $200\times 160\times 60~ (h_{min}=0.01D)$ &  0.921 & 89.2 \\
\end{tabular*}
{\rule{\temptablewidth}{1.0pt}}
\end{center}
\caption{\label{cylinder-cd-theta} Turbulent flow past a cylinder: comparison of the mean drag coefficient $C_d$ and separation angle $\theta_{\text{sep}}$ obtained by different schemes.}
\end{table}

\begin{figure}[!h]
\centering
\includegraphics[width=0.73\textwidth]{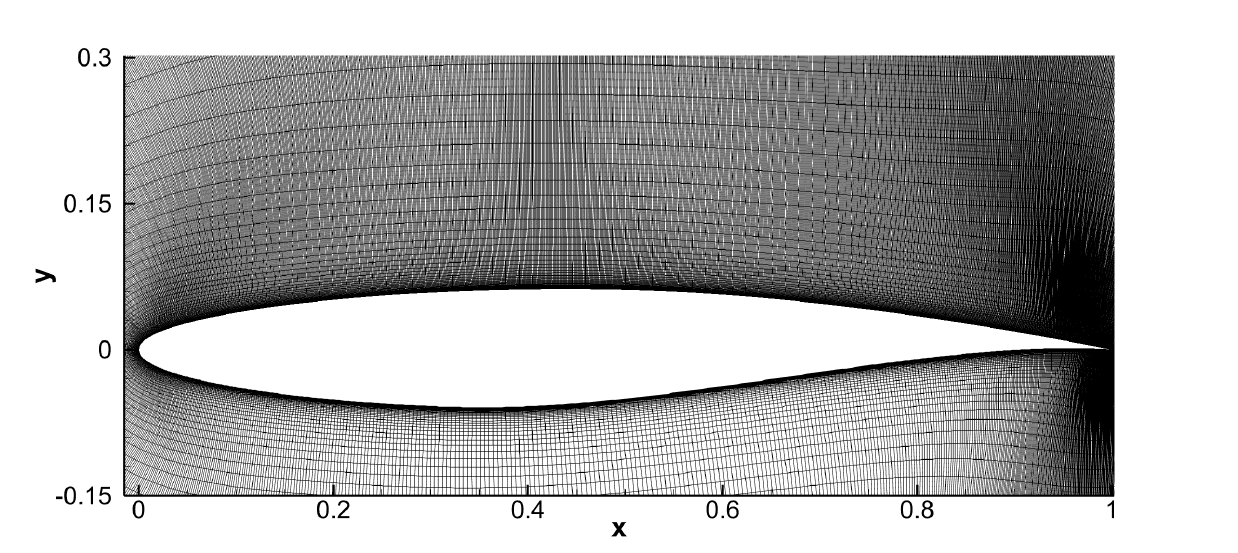}
\caption{\label{rae-mesh} Turbulent flow past an RAE 2822 airfoil: visualization of the computational mesh in the x-y plane.}
\end{figure}

\begin{figure}[!h]
\centering
\includegraphics[width=0.72\textwidth]{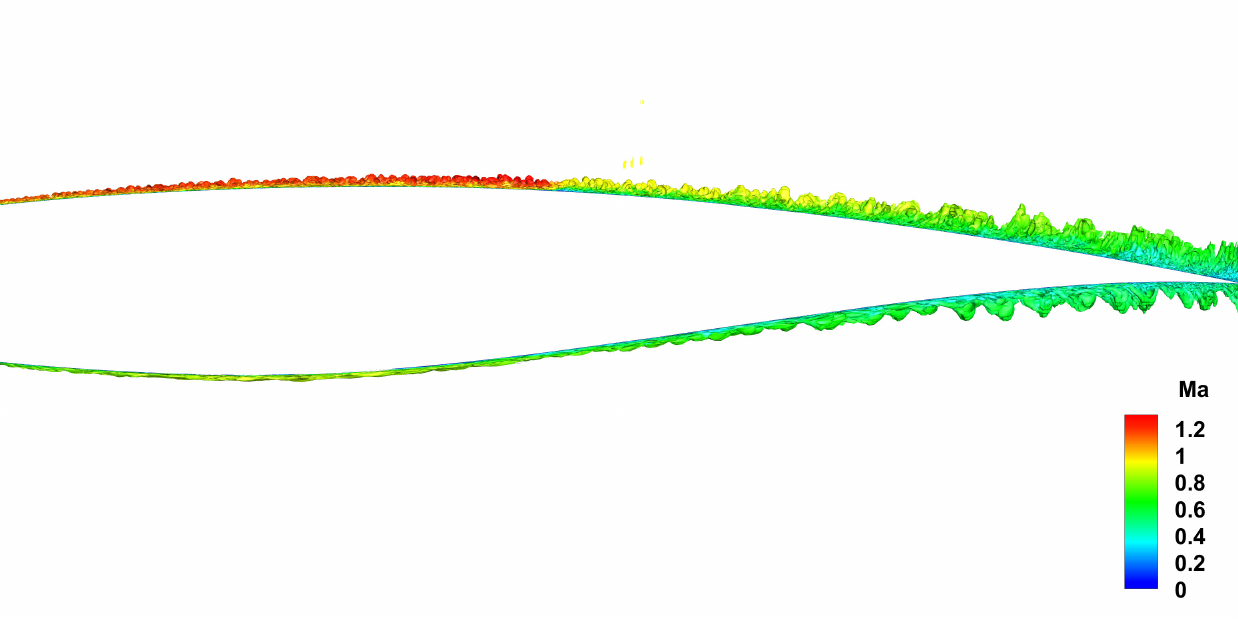}
\caption{\label{rae-3d-Q-1} Turbulent flow past an RAE 2822 airfoil: the iso-surface of the Q-criterion ($Q=3.0$) with a front view.}
\end{figure}

\begin{figure}[!h]
\centering
\includegraphics[width=0.6\textwidth]{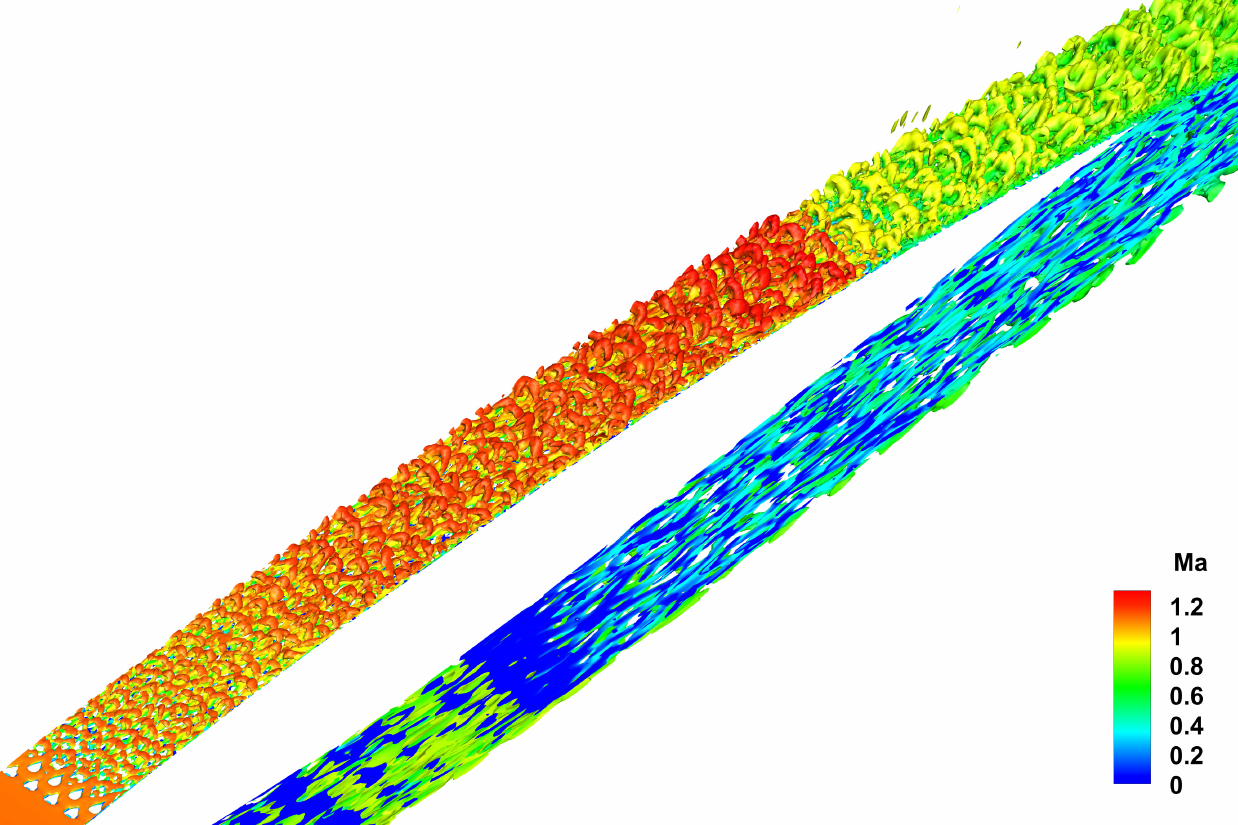}
\includegraphics[width=0.6\textwidth]{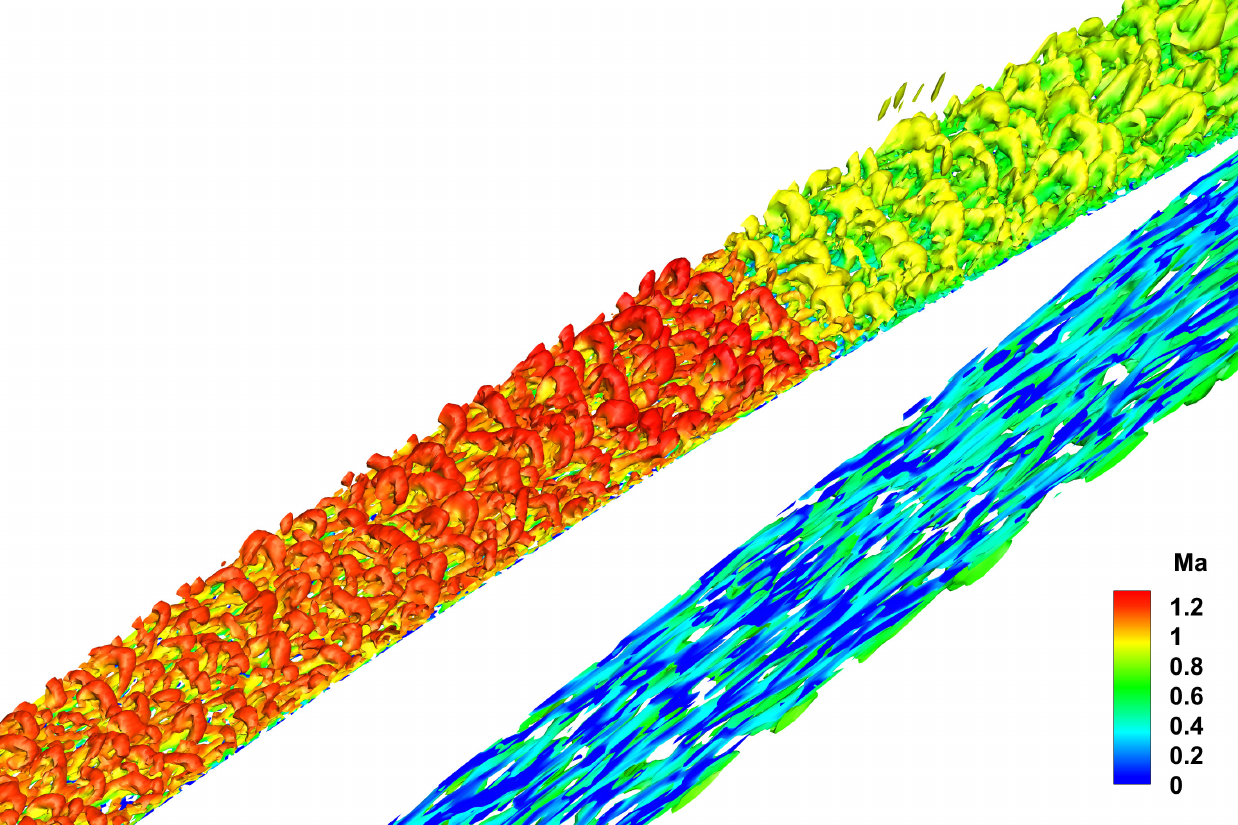}
\caption{\label{rae-3d-Q-2} Turbulent flow past an RAE 2822 airfoil: the iso-surface of the Q-criterion ($Q=3.0$) with local isometric views.}
\end{figure}

\begin{figure}[!h]
\centering
\includegraphics[width=0.495\textwidth]{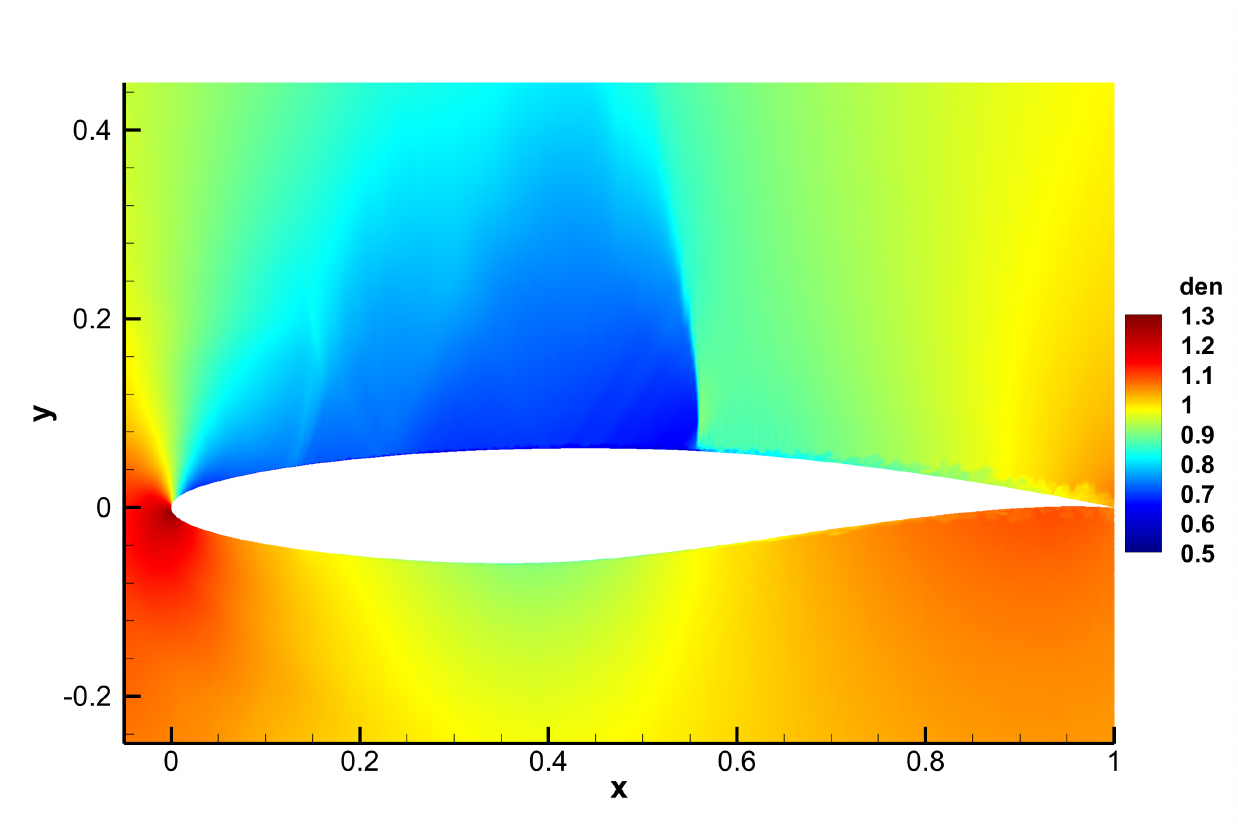}
\includegraphics[width=0.495\textwidth]{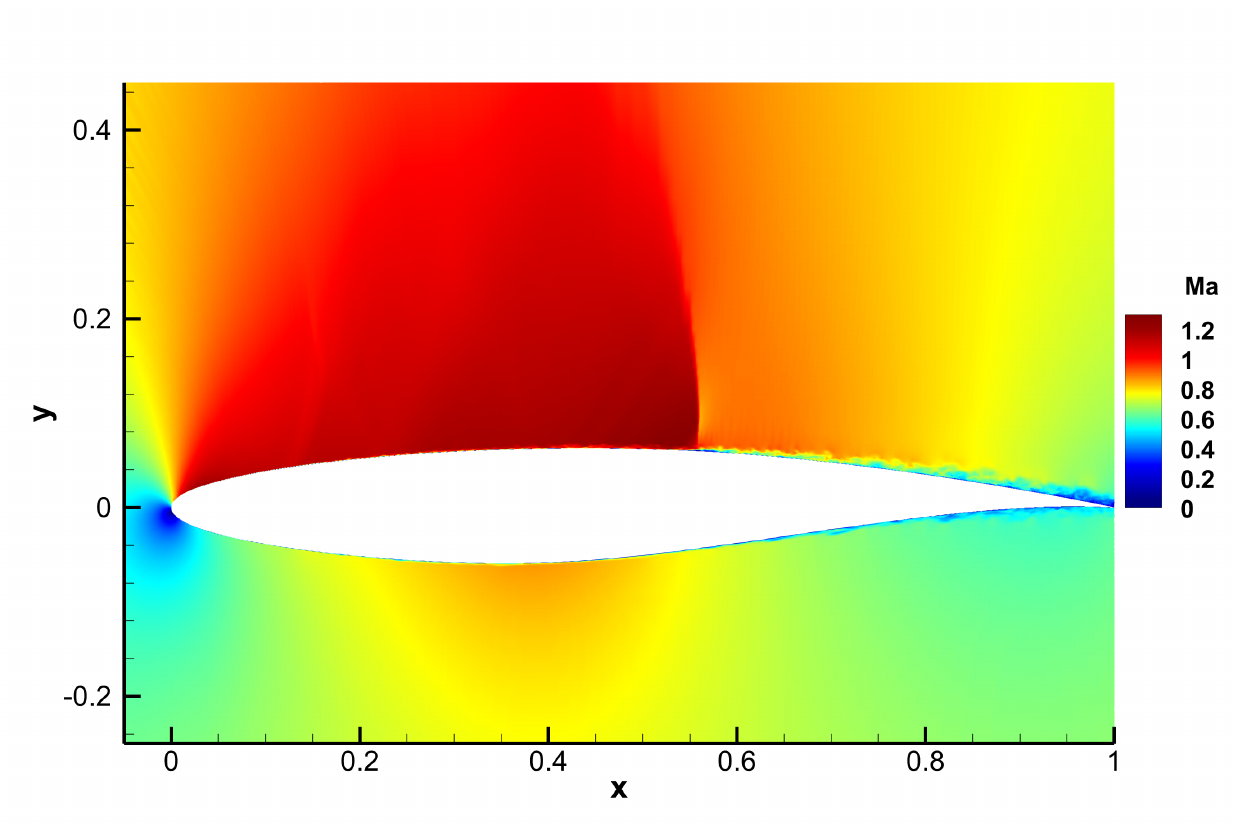}
\caption{\label{rae-2d-den-ma} Turbulent flow past an RAE 2822 airfoil: the instantaneous density contour and Mach number contour in the mid-span plane.}
\end{figure}

\begin{figure}[!h]
\centering
\includegraphics[width=0.495\textwidth]{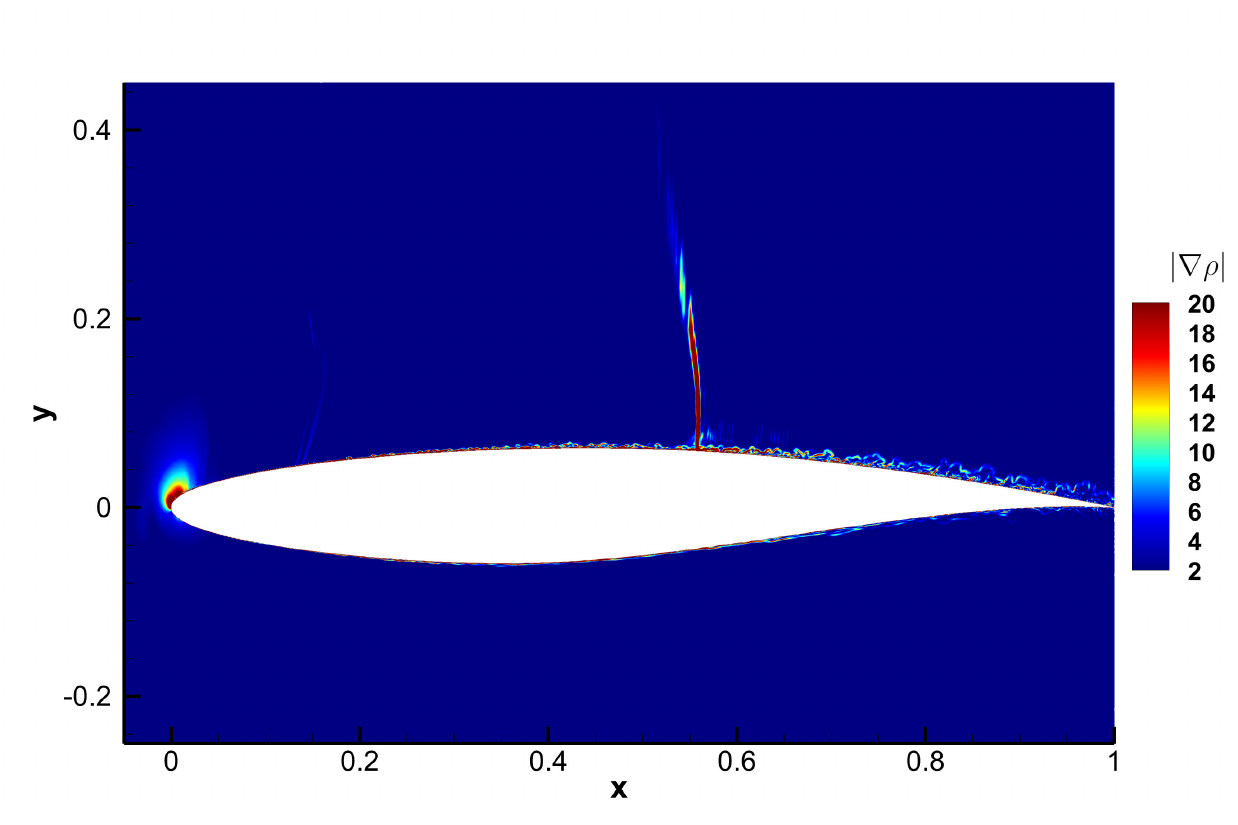}
\includegraphics[width=0.495\textwidth]{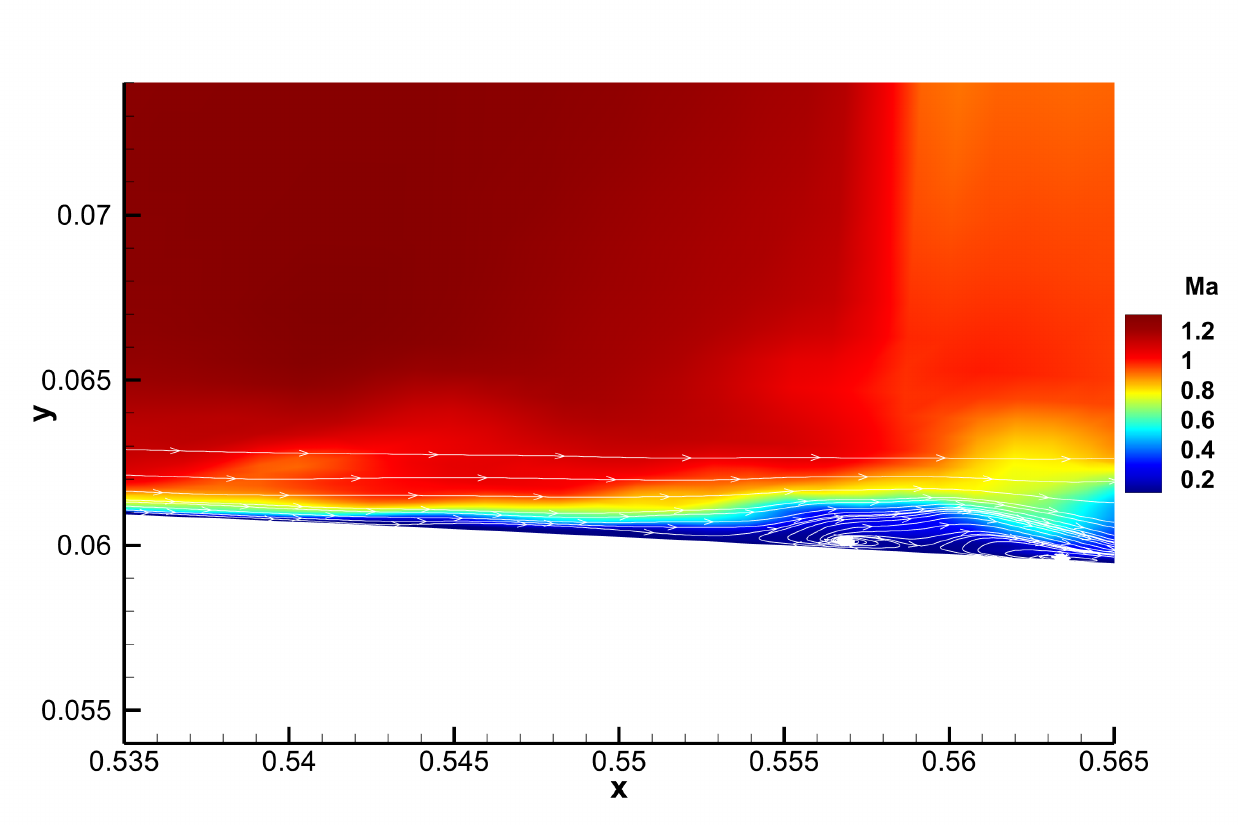}
\caption{\label{rae-2d-ma-local} Turbulent flow past an RAE 2822 airfoil: the instantaneous density gradient magnitude contour in the mid-span plane and Mach number contour near the shock region with in-plane streamlines.}
\end{figure}

\begin{figure}[!h]
\centering
\includegraphics[width=0.6\textwidth]{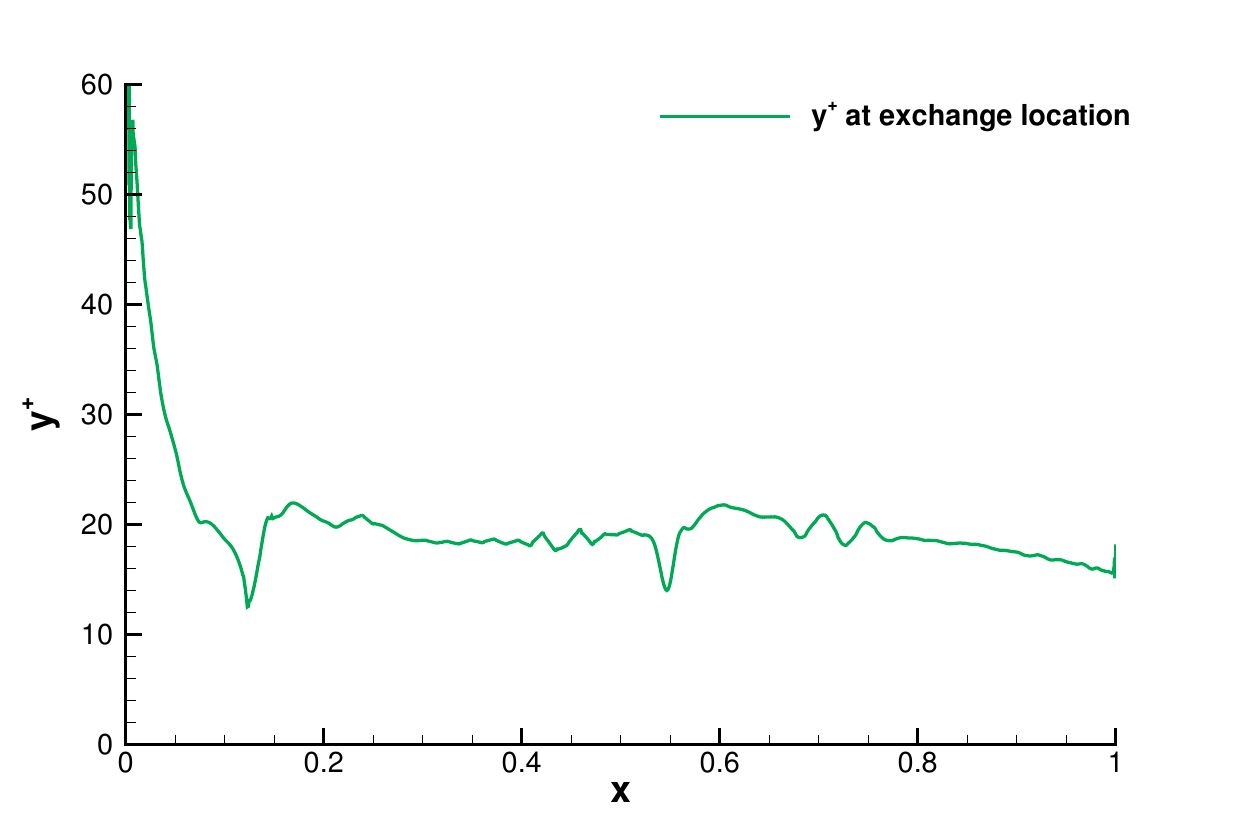}
\caption{\label{rae-yplus} Turbulent flow past an RAE 2822 airfoil: the time- and spanwise-averaged $y^+$ at the exchange location along the upper surface.}
\end{figure}

\begin{figure}[!h]
\centering
\includegraphics[width=0.495\textwidth]{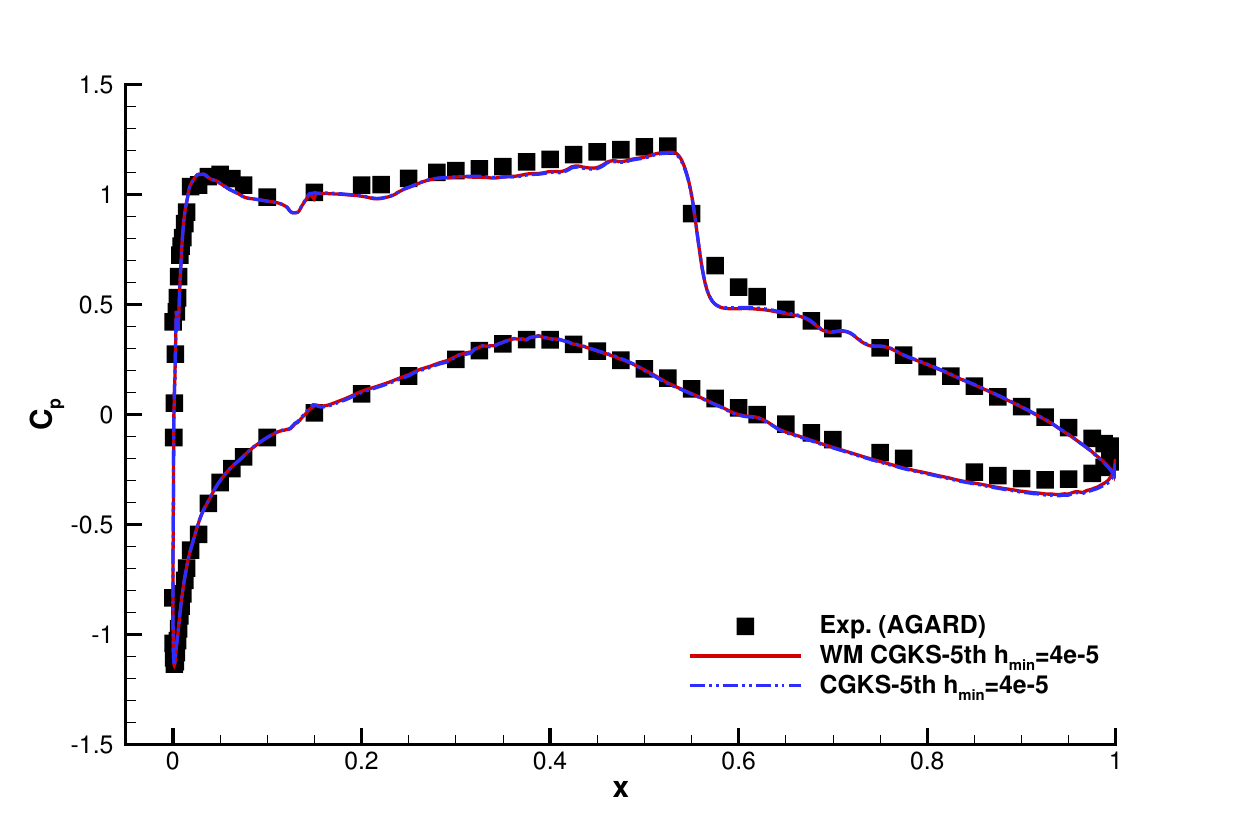}
\includegraphics[width=0.495\textwidth]{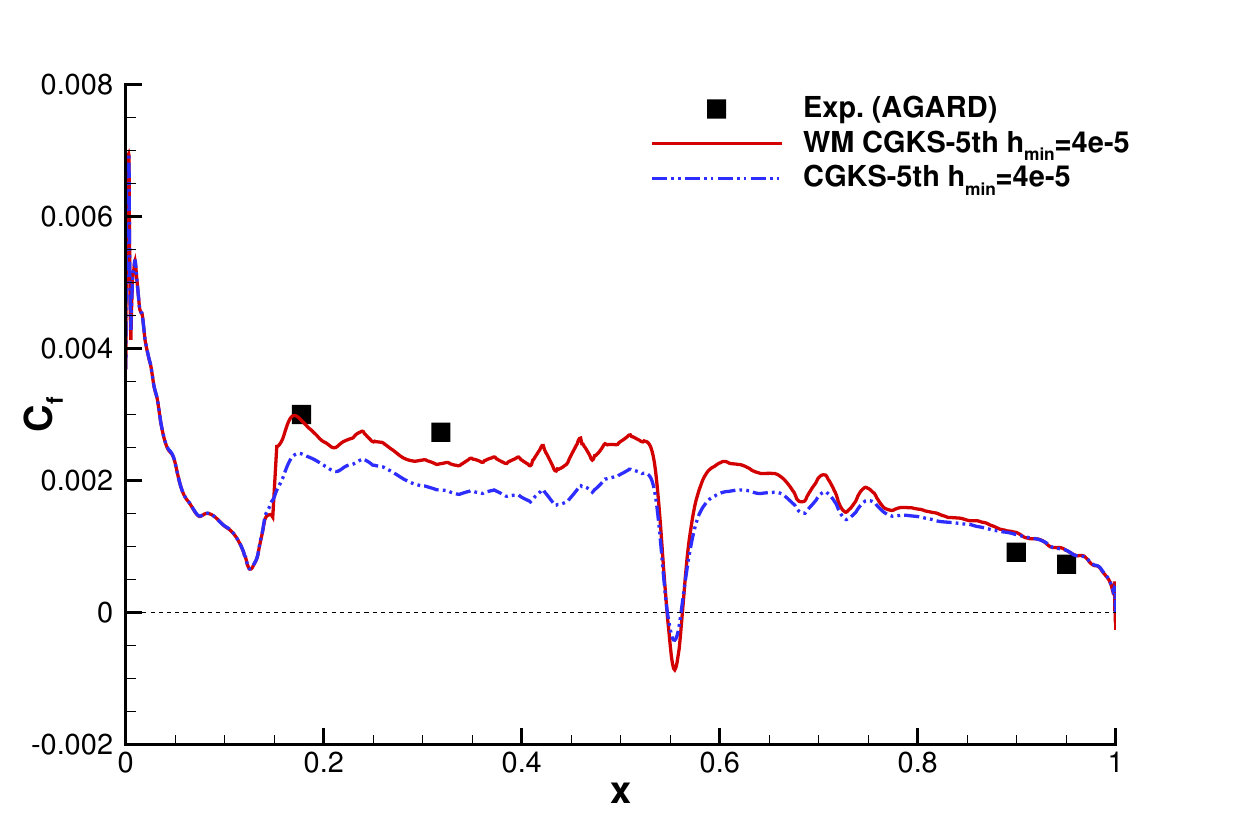}
\caption{\label{rae-2d-cp-cf} Turbulent flow past an RAE 2822 airfoil: the time- and spanwise-averaged pressure coefficient distribution along the airfoil surface (left) and skin friction coefficient distribution along the upper surface (right).}
\end{figure}

\begin{figure}[!h]
\centering
\includegraphics[width=0.62\textwidth]{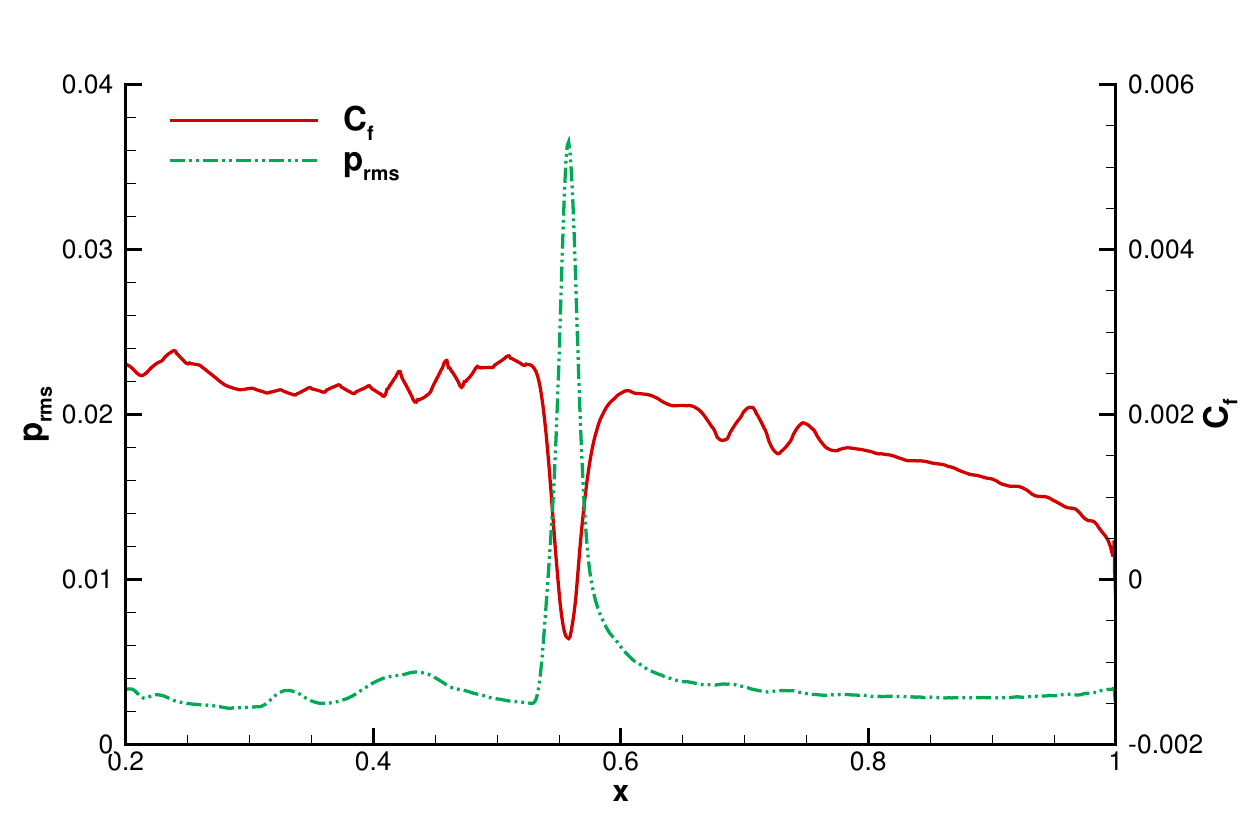}
\caption{\label{rae-prms-cf} Turbulent flow past an RAE 2822 airfoil: the time- and spanwise-averaged skin-friction coefficient $C_f$ (right axis) and wall-pressure fluctuation $p_{rms}$ (left axis) along the upper surface.}
\end{figure}

\subsection{Turbulent flow past an RAE 2822 airfoil}

To assess the capability of the proposed WM CGKS-5th in handling high-Reynolds-number wall-bounded flows with shock-wave/boundary-layer interaction (SWBLI), the transonic flow over an RAE 2822 supercritical airfoil is considered. This case features complex SWBLI with flow separation under stringent near-wall resolution requirements, constituting a challenging benchmark for the wall-modeling strategy.

The free-stream conditions correspond to Case 6 of the AGARD experiment~\cite{RAE2822-AGARD} with corrected parameters~\cite{RAE2822-EUROVAL, RAE2822-ref-1}: the free-stream Mach number is $Ma_\infty = 0.729$, the chord-based Reynolds number is $Re_c = 6.5 \times 10^6$, and the angle of attack is $\alpha = 2.31^\circ$.

Although the RAE 2822 geometry is two-dimensional, a three-dimensional computation is performed to resolve the spanwise turbulent structures.
The computational domain is extended in the spanwise direction with a periodicity length of $L_z = 0.03c$~\cite{RAE2822-ref-2}, where $c=1$ denotes the chord length.
The mesh consists of $1920 \times 80 \times 64$ cells in the wrap-around, wall-normal, and spanwise directions, respectively, clustered toward the airfoil surface with a minimum wall-normal spacing of $h_{\min} = 4 \times 10^{-5}c$.
A view of the mesh topology in the $x$--$y$ plane is shown in Figure~\ref{rae-mesh}.

On the airfoil surface, a no-slip isothermal wall boundary condition is imposed with the wall temperature set equal to the free-stream static temperature, $T_w = T_\infty$, and a far-field condition is applied at the outer boundary.
The flow field is initialized uniformly with the free-stream state $(\rho, U, V, W, p)_\infty = (1,\, Ma_\infty \cos\alpha,\, Ma_\infty \sin\alpha,\, 0,\, 1/\gamma)$, and the viscosity is evaluated using Sutherland's law.

To accelerate the development of the turbulent flow field, a two-dimensional simulation is first carried out in the $x$--$y$ plane with identical mesh distribution and flow conditions until a quasi-steady state is reached.
The resulting two-dimensional solution is then extruded uniformly in the spanwise direction to initialize the three-dimensional computation.
To promote transition to turbulence on the airfoil surface, a wall-normal blowing-and-suction disturbance is applied in the region $0.12 \leqslant x/c \leqslant 0.15$ on the upper surface, following~\cite{RAE2822-ref-2}, given by
\begin{equation}
  v_n(x,z,t) = A \sin(\pi\theta)\,\sin(2\pi\alpha_x\theta + 2\pi\beta t)\,
  \sin\!\left(\frac{2\pi\alpha_z z}{L_z}\right),
\end{equation}
where $\theta = (x - x_1)/(x_2 - x_1)$ with $x_1/c = 0.12$ and $x_2/c = 0.15$.
The disturbance amplitude is $A = 0.08$, the streamwise and spanwise wavenumber parameters are $\alpha_x = 2$ and $\alpha_z = 4$, respectively, and the temporal frequency parameter is $\beta = -1$.

Figure~\ref{rae-3d-Q-1} and~\ref{rae-3d-Q-2} present instantaneous isosurfaces of the Q-criterion ($Q = 3$) obtained by the WM CGKS-5th, colored by the local Mach number.
The flow development on the airfoil surface is clearly captured. Near the leading edge, the boundary layer remains laminar with no resolved vortical structures.
Downstream of the blowing-and-suction region ($0.12 \leqslant x/c \leqslant 0.15$), spanwise-coherent structures emerge and rapidly break down into fine-scale three-dimensional eddies, indicating a complete transition to turbulence well upstream of the shock.
The supersonic turbulent boundary layer ahead of the shock exhibits dense vortical structures with $Ma > 1$, while a sharp Mach number drop across the shock reflects the abrupt deceleration to subsonic speeds.
Downstream of the shock, the vortical structures broaden noticeably, consistent with boundary-layer thickening under the adverse pressure gradient.
These results show that the WM CGKS-5th captures the transition process and the post-shock boundary-layer development on the coarse wall-modeled mesh.

Figures~\ref{rae-2d-den-ma} and~\ref{rae-2d-ma-local} present instantaneous flow fields around the airfoil on the mid-span plane.
In the density and Mach number contours, the flow accelerates rapidly over the leading edge and forms an extensive supersonic zone, terminated by a well-defined shock near $x/c \approx 0.56$.
The density contour exhibits a sharp jump across the shock, indicating the shock-capturing capability of the CGKS-5th on the present coarse wall-modeled mesh. This observation is further corroborated by the density-gradient magnitude field shown in Figure~\ref{rae-2d-ma-local} (left), where the shock front appears as a thin, crisp band with peak $|\nabla\rho|$ values sharply concentrated.
The close-up view of the shock region (Figure~\ref{rae-2d-ma-local}, right), overlaid with in-plane streamlines, reveals that the boundary layer thickens appreciably upstream of the shock foot and that the streamlines deflect markedly away from the wall immediately downstream, indicating a shock-induced separation driven by the strong adverse pressure gradient.
Together, these instantaneous results show that the WM CGKS-5th resolves the shock, the induced separation, and the post-shock boundary-layer response on a relatively coarse near-wall grid.

Figure~\ref{rae-yplus} shows the time- and spanwise-averaged $y^+$ at the exchange location along the upper surface.
Except for a thin region near the leading edge, where the boundary layer is laminar and extremely thin, the exchange-location $y^+$ remains within approximately $18$--$22$ over the entire chord.
Since the wall model integrates the near-wall momentum balance directly on the embedded mesh rather than fitting a logarithmic law, this level keeps the modeled layer thin while still spanning the inner layer, and the wall model therefore operates in the first regime described in Section~\ref{sec:wall_model}, acting as a turbulence closure for the unresolved inner layer.

Figure~\ref{rae-2d-cp-cf} compares the time- and spanwise-averaged pressure coefficient $C_p$ and skin-friction coefficient $C_f$ on the upper surface obtained with and without the wall model on the same coarse mesh.
In the $C_p$ distribution, both computations reproduce the suction-side pressure plateau in the supersonic region and capture the shock location in close agreement with the experiment.
The difference between the two approaches becomes more pronounced in the $C_f$ distribution.
Without the wall model, the CGKS-5th systematically under-predicts the skin friction over the entire upper surface, a characteristic consequence of insufficient near-wall resolution on this coarse mesh.
In contrast, the WM CGKS-5th returns appreciably higher $C_f$ levels that are markedly closer to the experimental data.
For instance, near the transition region at $x/c \approx 0.2$, the WM CGKS-5th predicts $C_f \approx 0.003$, in close agreement with the measured value, whereas the CGKS-5th yields only $C_f \approx 0.0024$.
Near the shock foot, both solutions exhibit a sharp drop in $C_f$ crossing zero, consistent with the shock-induced incipient separation observed in the instantaneous flow field.
Downstream of the shock, the WM result recovers to a positive friction level more rapidly.
These results confirm that the wall model is essential for obtaining quantitatively reliable wall shear stress on the present coarse mesh, whereas the pressure field is comparatively less sensitive to the near-wall treatment.

To further characterize the SWBLI, Figure~\ref{rae-prms-cf} shows the time- and spanwise-averaged skin-friction
coefficient $C_f$ together with the wall-pressure fluctuation $p_{rms}$ along the upper surface.
At $x/c \approx 0.56$, $C_f$ drops sharply to a near-zero minimum, indicating the shock and its induced incipient separation, before recovering downstream.
At the same location, $p_{rms}$, which remains low and nearly flat over the rest of the chord, exhibits a pronounced peak.
The coincidence of the $C_f$ minimum and the $p_{rms}$ peak confirms that the strong wall-pressure fluctuations originate from the unsteady shock motion and the associated separation.

Table~\ref{rae-cd-cl} lists the mean drag and lift coefficients.
Both computations on the coarse mesh yield reasonable predictions of the integral aerodynamic forces.
Since $C_l$ is predominantly determined by the surface pressure distribution, which is nearly identical for the two schemes, the resulting lift coefficients are expectedly close, their difference being within the statistical and grid uncertainty.
For the drag coefficient, wave drag dominates over friction drag in transonic flow, so the pronounced difference in $C_f$ between the two approaches does not translate into an appreciable difference in $C_d$.
The benefit of the wall model is therefore primarily reflected in the local wall shear stress prediction.

\begin{table}[!h]
\begin{center}
\def\temptablewidth{0.88\textwidth}{\rule{\temptablewidth}{1.0pt}}
\begin{tabular*}{\temptablewidth}{@{\extracolsep{\fill}}c|c|c|c} 
Scheme & Mesh  & $C_d$ & $C_l$   \\
\hline
Exp.(AGARD) & - & 0.0128 & 0.743  \\
WM CGKS-5th & $1920\times 80\times 64~ (h_{min}=4 \times 10^{-5})$ & 0.0116  & 0.732\\
CGKS-5th & $1920\times 80\times 64~ (h_{min}=4 \times 10^{-5})$ & 0.0114  & 0.735 \\
\end{tabular*}
{\rule{\temptablewidth}{1.0pt}}
\end{center}
\caption{\label{rae-cd-cl} Turbulent flow past an RAE 2822 airfoil: comparison of the mean drag coefficient and lift coefficient obtained by different schemes.}
\end{table}

\subsection{Computational efficiency}

In this subsection, the computational efficiency of the proposed WM CGKS-5th is evaluated. To quantify the extra overhead introduced by the wall model, Table~\ref{efficiency} compares the computational cost of the CGKS-5th and the WM CGKS-5th per 1000 time steps, measured on the respective computational meshes of the two cases in the same multi-GPU environment. For both the cylinder and the RAE 2822 cases, the cost of the WM CGKS-5th is nearly identical to that of the baseline CGKS-5th, with a marginal increase of only $0.5\%$ to $0.7\%$.

This negligible overhead is expected, since the additional wall-modeling procedures are strictly confined to the first layer of off-wall cells, which constitute a minuscule fraction of the total three-dimensional computational domain. 
Given the nearly identical per-step cost, the principal advantage of the WM CGKS-5th lies in its grid-reduction capability.
By sustaining high accuracy on meshes far coarser than those required by a wall-resolved computation, the proposed scheme avoids the prohibitive cell counts that arise from resolving the near-wall region at high Reynolds numbers.
To put this coarsening in perspective, the near-wall mesh of the high-Reynolds-number RAE 2822 case corresponds to an exchange-location $y^+\approx20$, whereas a wall-resolved computation requires a first-cell $y^+\approx1$. Recovering wall-resolved resolution would therefore demand a roughly twentyfold wall-normal refinement near the wall, together with a comparable streamwise and spanwise refinement to resolve the near-wall turbulent scales; the resulting growth in cell count, and the attendant time and memory cost, becomes increasingly prohibitive as the Reynolds number rises.
This translates into substantial savings in both computational time and memory, making the WM CGKS-5th attractive for engineering-scale simulations.

\begin{table}[!h]
\begin{center}
\def\temptablewidth{0.88\textwidth}{\rule{\temptablewidth}{1.0pt}}
\begin{tabular*}{\temptablewidth}{@{\extracolsep{\fill}}c|c|c|c|c} 
Case & Scheme & Mesh & Time (s) & Ratio  \\
\hline
Cylinder & CGKS-5th & $224\times 160\times 60$ & 827.3 & 1 \\
Cylinder & WM CGKS-5th & $224\times 160\times 60$ & 831.6 & 1.005 \\
\hline
\hline
RAE 2822 & CGKS-5th & $1920\times 80\times 64$ & 6936.3  & 1 \\
RAE 2822 & WM CGKS-5th & $1920\times 80\times 64$ & 6986.5  & 1.007
\end{tabular*}
{\rule{\temptablewidth}{1.0pt}}
\end{center}
\caption{\label{efficiency} Computational efficiency: computational time per 1000 time steps for the CGKS-5th and the WM CGKS-5th.}
\end{table}

\section{Conclusion}

This work extends wall-modeled simulation methodology to the gas-kinetic scheme framework by integrating a non-equilibrium wall model with a recently developed fifth-order compact gas-kinetic scheme (CGKS-5th), yielding an affordable framework for high-Reynolds-number wall-bounded turbulent flows.
The wall model retains the wall-parallel pressure-gradient source term and employs a pressure-gradient-corrected generalization of the van Driest damping function, enabling it to account for departures from local equilibrium in adverse-pressure-gradient and separated regions.
The coupling is realized purely as a boundary-condition correction, requiring no modification to the underlying CGKS-5th solver and incurring only negligible overhead, $0.5\%$--$0.7\%$ per time step.
The entire framework is realized in a multi-GPU implementation parallelized via CUDA and MPI, which makes these high-fidelity separated-flow simulations tractable on a modest multi-GPU workstation.
The framework has been validated using two cases with distinct separation mechanisms.
The circular-cylinder case assesses geometry-induced laminar separation and vortex shedding in a turbulent wake, while the RAE 2822 airfoil case examines SWBLI and shock-induced incipient separation.
In both cases, the wall-modeled CGKS-5th achieves improved predictions on coarse near-wall meshes.
In particular, it improves the skin-friction peak and base pressure for the circular cylinder, and captures the shock structure, incipient separation, and upper-surface skin-friction distribution for the RAE 2822 airfoil.
The two cases also exercise the wall model in its two operating regimes: pressure-gradient-corrected near-wall reconstruction in the laminar cylinder boundary layer, and inner-layer turbulence closure in the attached turbulent boundary layer of the airfoil.
These results demonstrate that the proposed coupling provides an accurate and low-overhead route to wall-modeled simulation within the CGKS framework.
Future work will further refine the wall-model formulation and extend the framework to more complex three-dimensional and engineering-relevant configurations.

\section*{Acknowledgements}

The current research is supported by National Science Foundation of China (92371107), National Key R$\&$D Program of China (Grant Nos.
2022YFA1004500), and Hong Kong research grant council (16208324).

\section*{Declaration of competing interest}

The authors declare that they have no known competing financial interests or personal relationships 
that could have appeared to influence the work reported in this paper.

\section*{Data availability}

The data that support the findings of this study are available from
the corresponding author upon reasonable request.

\end{document}